\documentclass[useAMS,usenatbib]{mn2e}
\usepackage{comment}
\usepackage{graphicx}
\usepackage{color}
\usepackage{subfigure}

\newcommand{\shlA}[0]{\color{black}}   
\newcommand{\ehlA}[0]{\color{black}} 

\newcommand{\shl}[0]{\color{black}}   
\newcommand{\ehl}[0]{\color{black}} 
\newcommand{\hl}[1]{\color{black}#1\color{black}\ } 
\newcommand{\sbl}[0]{} 

\title[The Water Maser in MG 0414+0534: The Influence of Gravitational Microlensing]{The Water Maser in MG 0414+0534: The Influence of Gravitational Microlensing}
\author[H. Garsden et al.]{H. Garsden$^{1}$\thanks{E-mail:
hgar7294@uni.sydney.edu.au  (HG); geraint.lewis@sydney.edu.au (GFL); lisa.harvey-smith@csiro.au (LHS)}, 
G. F. Lewis$^{1}$\footnotemark[1] and L. Harvey-Smith$^{1,2}$\footnotemark[1]\thanks{Research undertaken as part of the Commonwealth Cosmology 
Initiative (CCI: www.thecci.org), an international collaboration 
supported by the Australian Research Council.}\\
$^{1}$Sydney Institute for Astronomy, School of Physics, A28, The University of Sydney, NSW 2006, Australia\\
$^2$CSIRO Astronomy and Space Science, Australia Telescope National Facility, PO Box 76, Epping, NSW, 2121, Australia
}
\begin{document}
\date{Accepted 2010 December 17}

\pagerange{\pageref{firstpage}--\pageref{lastpage}} \pubyear{2010}

\maketitle

\label{firstpage}

\begin{abstract}
Water masers have been observed in several high redshift active galactic nuclei, including the 
gravitationally lensed quasar
MG 0414+0534. This quasar is lensed into four images, and the water maser is detected in two of them. The 
broadening of the maser emission line and its velocity offset are consistent with a group of masers 
associated with a quasar jet.
If the maser group is microlensed we can probe its structure and size  by observing its microlensing behaviour over time. We present  results of a high resolution numerical analysis  of microlensing of the  maser in MG 0414+0534,
using several physically motivated maser models covering a range of sizes and emission profiles.  Time-varying  spectra of the microlensed maser  are generated, displayed, and analysed, and the behaviour of the different models compared. The observed maser line in MG 0414+0534 is consistent
with maser spots as in other quasar jets, provided substructure is de-magnified  or currently lost in noise; otherwise
 smooth extended maser models  are also candidates to generate the observed spectrum.
Using measures of spectral variability  we find that if the maser has small
substructure of $\sim 0.002$ pc then a variation of \shl  0.12 mag in flux and
2.0 km s$^{-1}$ in velocity centroid \ehl of the maser line could be observed within 2 decades. For the smallest maser model in this study a magnification of $>$ 35  is possible 22\% of the time, which is of significance in the search for other lensed masers. 
\end{abstract}


\begin{figure*}
\centering
\includegraphics[scale=0.7]{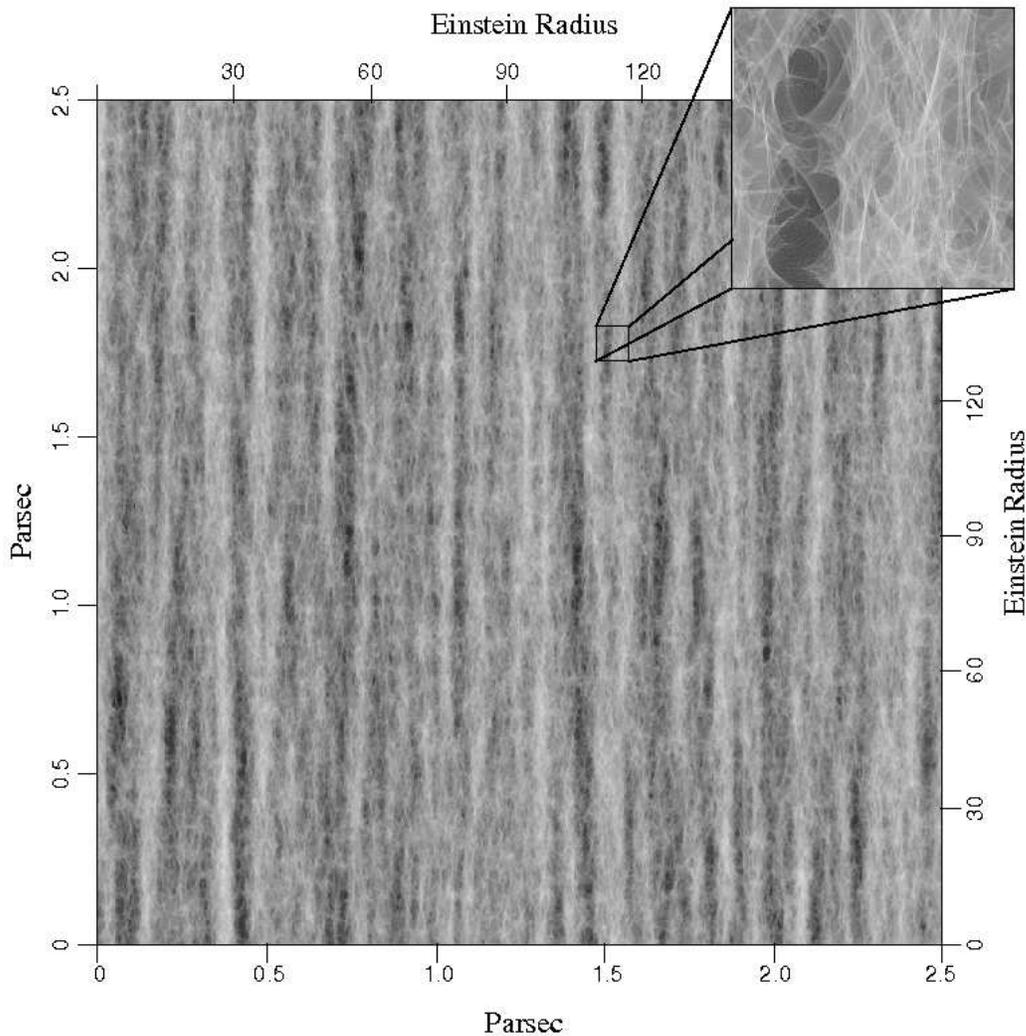}
\caption{The magnification map representing microlensing variability in image A1, one of the four images of the  quasar produced by lensing in MG 0414+0534. 
The map covers a region of the source plane;  at all locations the brightness indicates
how much a pixel-sized source would be (de)magnified if the source were at that location. Bright areas indicate
high magnification.  There are  6,902,710  M$_\odot$  compact objects in the lens.  The map size is   $2.5 \times 2.5$ pc$^2$ ($185 \times 185$ ER$^2$) with a resolution of 26 AU. 
The
break-out box is 0.1$\times$0.1 pc$^2$ (\hl{7.4 $\times$ 7.4 ER$^2$}) at the same resolution. The break-out box shows the high density of caustics
in the source plane of this system, and that there are patches where there are no caustics and patches
where there are many; these appear as light and dark regions on the larger map. \sbl The 
microlensing parameters used to generate the map are $\sigma$ (convergence) = 0.472, and $\gamma$ (shear) = 0.478 (see Section \ref{analysis_described}).\ehl}
\label{big_map}
\end{figure*}

\begin{table*}
 \centering
 \begin{minipage}{120mm}
 \caption{Properties of the magnification maps used.}
 \label{parameters}
 \begin{tabular}{@{}ccccccc}
  \hline
  Image & $\kappa$ & $\gamma$ &  \hl{Sidelength}  & Resolution & Number of lens objects ($M_{\odot}$)\\
        &          &          &       &         & \\
   \hline
   A1 & 0.472 & 0.478 &  2.5  pc \hl{(185  ER)} & 26 AU  & 3,204,158    \\
      &       &       & 15  pc \hl{(1115 ER)} & 155 AU & 115,353,274   \\
   A2 & 0.485 & 0.550 &  2.5 pc \hl{(185 ER)} & 26 AU  & 6,710,081    \\
      &       &       & 15  pc  \hl{(1115 ER)} & 155 AU & 241,562,900   \\
  \hline
 \end{tabular}

\sbl

 Four magnification maps were used in this study. Images A1 and A2 were modelled, but there
are two maps of different resolutions for each image  because there are a range of sizes for the source models. 
$\kappa$  (convergence) and $\gamma$ (shear) are mass parameters from lens models for this system, $\kappa$ specifies the effect of
mass along a light path, and  $\gamma$  specifies the effects of  surrounding mass. 
 \ehl
\end{minipage}
\end{table*}

\section{Introduction}

Gravitational lensing  occurs when light travelling to Earth from distant sources is deflected
by the gravitational influence of an intervening massive object \citep{schneider,wambsganss_lr}.
The light source may be within the Galaxy and lensed by planetary systems, which is
useful in the search
for planets \citep[e.g.][]{gaudi}; or it may be from a very distant galaxy and lensed by another galaxy \citep[e.g.][]{willis} or 
galaxy cluster \citep[e.g.][]{sand}.
 In this paper we are concerned
with a quasar being lensed by a galaxy, many instances of which have been discovered \citep[e.g.][]{walsh,huchra,turner,myers,inada,kayo}.
In such systems, multiple magnified images of the quasar are produced, with the image properties  determined by the source emission profile, the lensing galaxy's  mass
distribution,  and the relative position of the source and lens.  Assuming a smooth mass distribution
for the lensing galaxy is
sometimes enough to model
the observed image configurations, but in other cases
the effect of the compact objects  within the  galaxy (stars, planets, etc.)
must also be considered.
Due to the intricate nature of the light paths through a galaxy of myriad objects, a slight change in the location of the
background source can produce a  change in the image magnifications (independent of any intrinsic
source variability). This effect of the compact structure is called \emph{microlensing} \citep{chang,young,paczynski,wambsganss_thesis,wambsganss_saas}, and  identifiable microlensing fluctuations of high amplitude are called \emph{high magnification events}. These events  can
last for relatively short durations -- months or even weeks \citep{shalyapin} -- and  usually occur with a frequency of decades.

The angular size of the source is important for  microlensing,
as larger sources will 
``wash  out'' the location-induced variability in magnification  \citep{lewis1,bate}, so it is small sources that produce the
most significant high magnification events. \shlA The shape of the source emission profile also affects microlensing
\citep{mortonson}. \ehlA
Therefore,  observations and modelling of  flux variability can provide an estimate of source size and flux emission profile. Apparent chromatic effects are also  seen 
 \hl{\citep{lewis2, eigenbrod}} because   if an extended source  emits different frequencies from separate regions, the frequencies will be magnified differently, and the source spectrum will be altered. Like the magnification, the spectrum may also change over
time, and studying spectral variability  can lead to possible spectral  emission profiles.  In these (and other) ways microlensing becomes a useful
tool for astronomical research \citep{Kayser,wambsganss_anp,kochanek,wambsganssr}.

\begin{figure*}
\centering    
\subfigure[Grainy Patch Schematic]{\includegraphics[width=55.3mm]{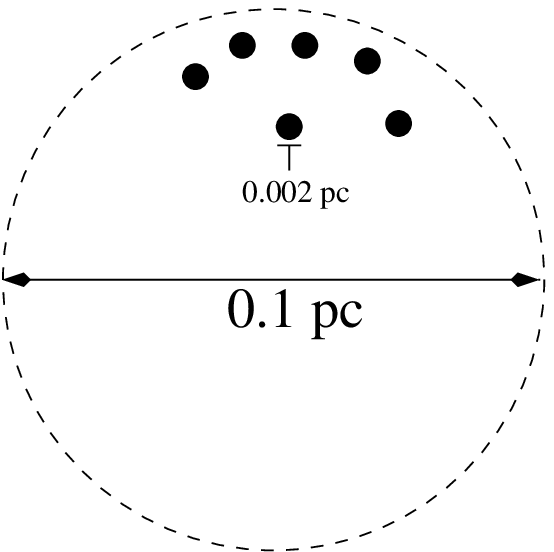}}
\subfigure[Grainy Ring Schematic]{\includegraphics[width=55.3mm]{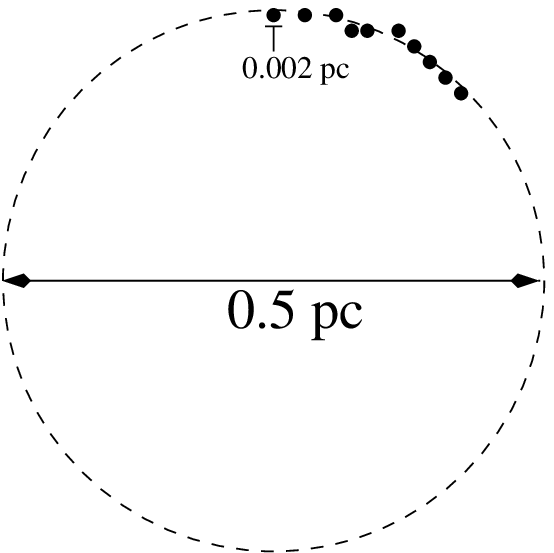}}
\subfigure[Hollow \& Solid Jets Schematic (1 jet shown)]{\includegraphics[width=57.41mm]{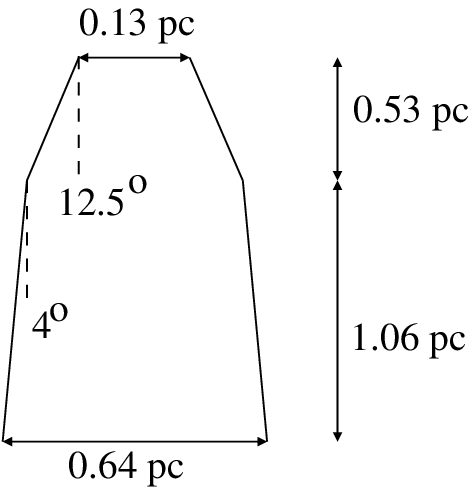}}
\subfigure[Grainy Patch: a patch of maser spots]{\includegraphics[width=55.3mm]{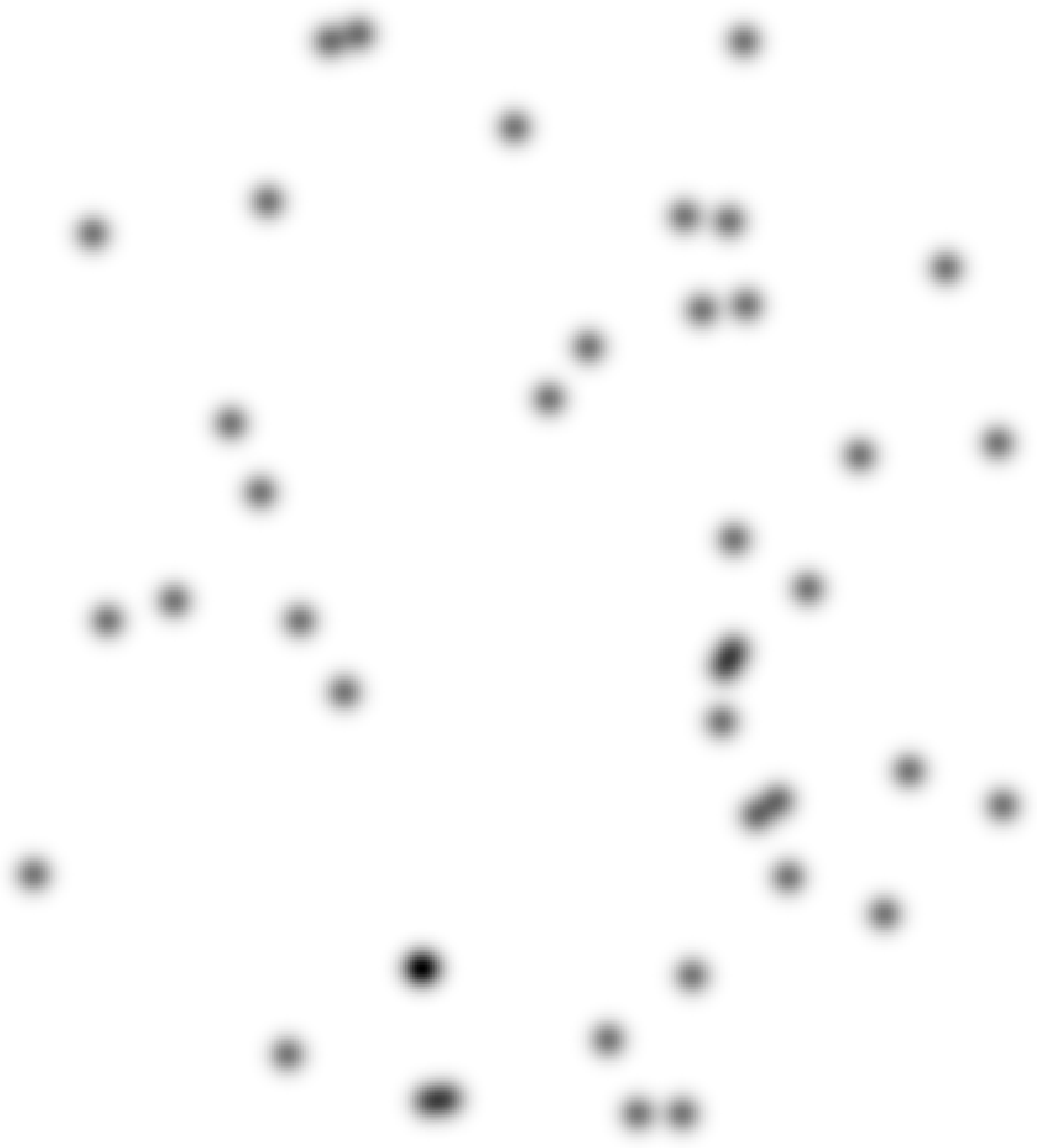}}
\subfigure[Grainy Ring: a ring of maser spots that can be oriented]{\includegraphics[width=55.3mm]{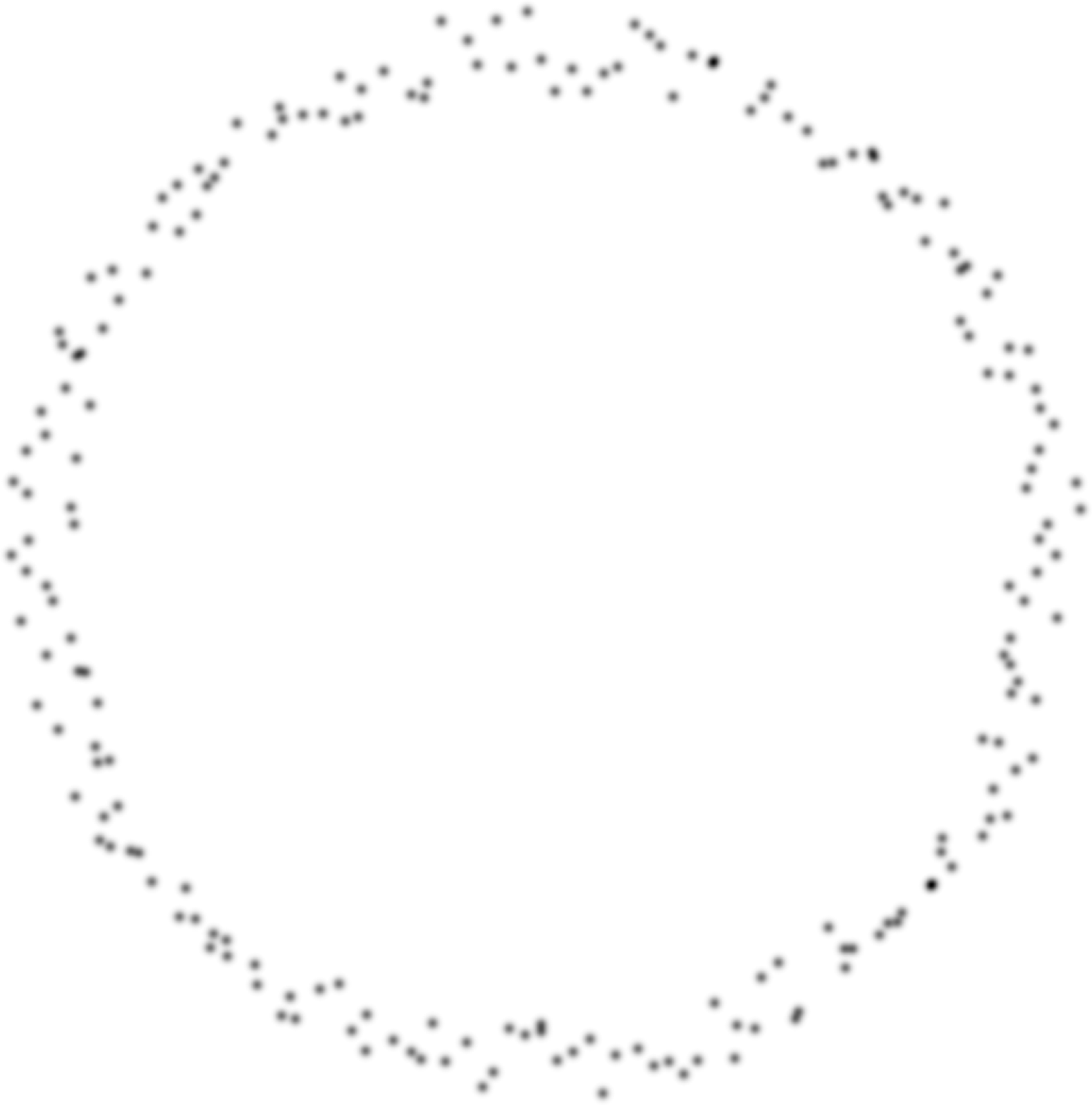}}
\subfigure[Hollow Jets: a bi-conical jet structure of smooth masers that can be oriented. The Solid Jet is the same but filled with maser material.]{\includegraphics[width=57.41mm]{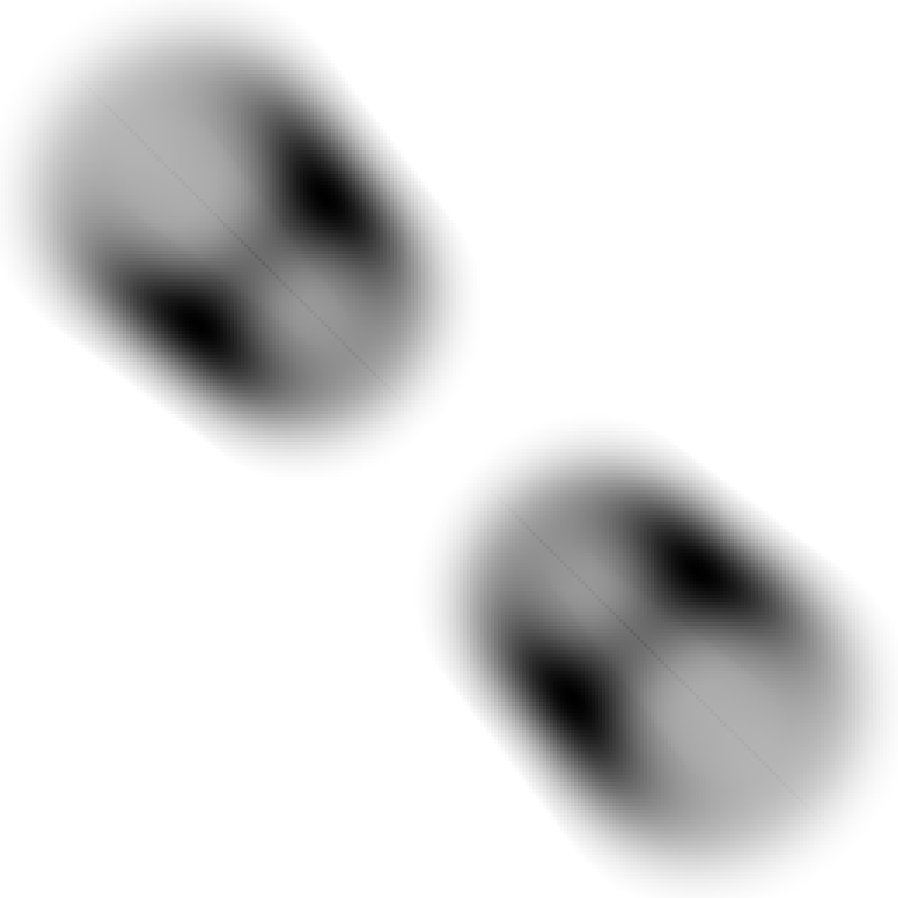}}
\caption{Schematics (a-c) and example images (d-f) of the three complex maser models (the Single Spot and
Smooth Patch are not shown). Figures (a) and (d)
show the \shl Grainy Patch;\ehl\ (b) and (e) the \shl Grainy Ring; \ehl (c) and (f) an example of  the Hollow Jets, which is bi-conical as in (f), but since  the jets are mirror images of each other only one schematic is shown in (c). The inclination of the jets in (f) is $i = 30\,^{\circ}$ South-East, relative to the magnification map as described in the text. The elements of the schematics are not to scale.}
\label{schemes}
\end{figure*}

In this paper we present results of numerical analyses of the microlensing of a water maser in a known microlensing system,
MG 0414+0534. We use several physically-motivated   models  to represent the maser source;  each model is characterised
by its shape, size,  emission and velocity profile.
 A  model
 is then subjected to numerical microlensing so that the alteration in the spectrum and flux after lensing can be found. 
The structure of the paper is as follows:
Section \ref{Background} discusses the maser in MG 0414+0534, and mechanisms used for   numerical microlensing analysis. In Section 
\ref{Method} we describe the parameters characterizing the MG 0414+0534 lens, we specify the maser source models, and explain how the models are used. Section \ref{Results}
presents the results of microlensing of the maser by examining which models induce the most variability,
what time scales may be expected for this, what forms the spectra take and how they vary over time, and whether the models match the current observation. 

\section{Background}
\label{Background}

\begin{figure*}
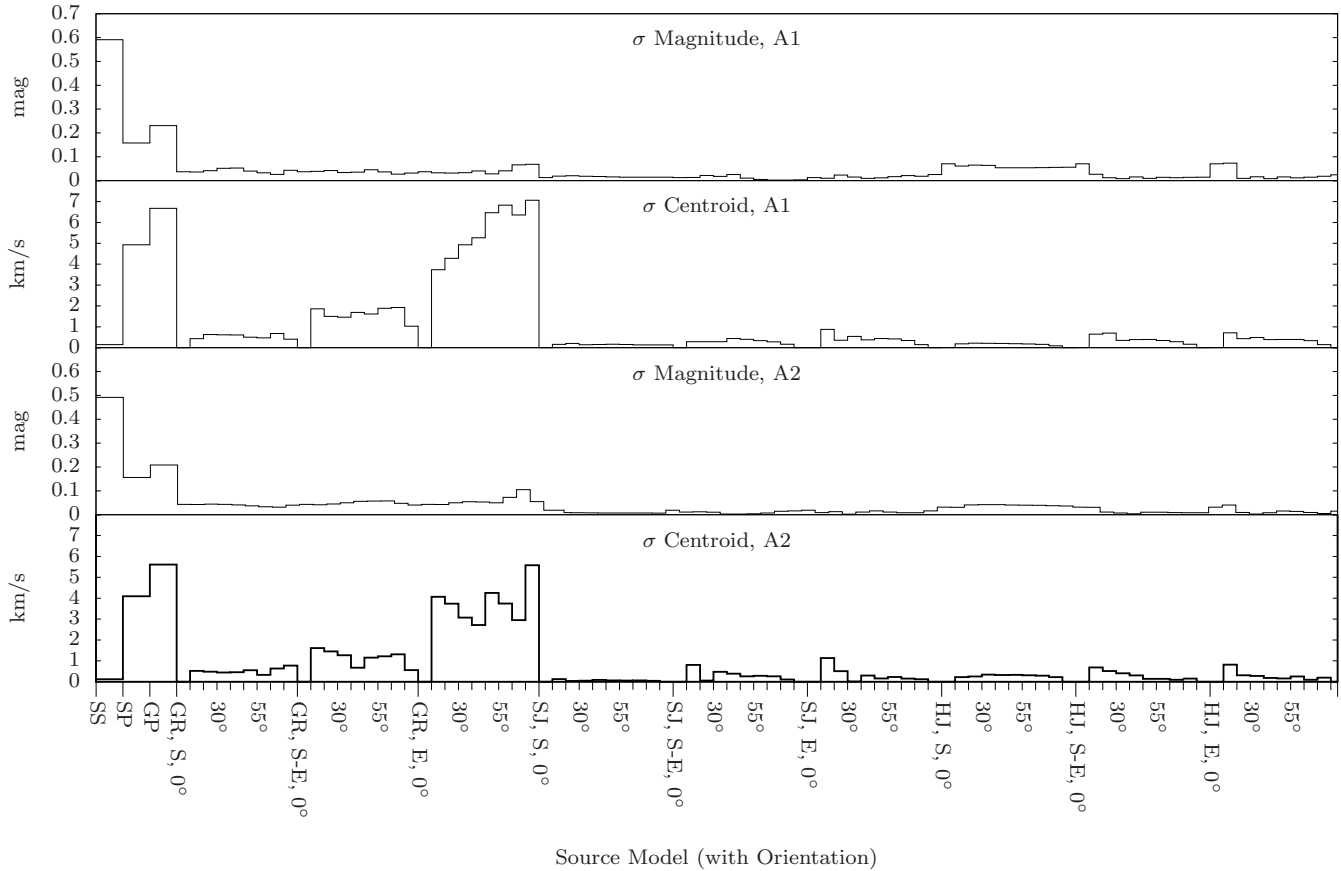

\centering
\begingroup%
\makeatletter%
\newcommand{\GNUPLOTspecial}{%
  \@sanitize\catcode`\%=14\relax\special}%
\setlength{\unitlength}{0.0500bp}%
%
\endgroup
\vspace{28mm}
\caption{\shl Measures of  variability  for the maser line in images A1 and A2, for all models,  after microlensing. 
The top panel is the
variation in the magnitude for image A1, the second panel  is the variation in the spectrum velocity centroid for image A1.  The next two panels are the same measures for image A2. In each panel is a plot of  variability measure (vertical axis) as a function of source model, and orientation, for those that can be oriented (horizontal axis).
Major tick marks indicate the model, inclination direction, and inclination angle e.g. GR=Grainy Ring,  S=South,  $0\,^\circ$ is the angle.
Models that cannot be oriented have no direction or angle.
Following a major tick mark, minor tick marks indicate increasing inclinations  for that model (not all angles shown).
The same horizontal axis applies to all panels and is only listed on the bottom one. The vertical axis is indicated
appropriately for each panel separately.
}
\label{overall}
\end{figure*}

\subsection{The lens and water maser MG 0414+0534}
\label{lens and maser}
MG 0414+0534   \citep{lawrence}  is a gravitational lens displaying  microlensing \citep{witt}. It
consists of a quasar at z = 2.639   being lensed
by a  galaxy at z = 0.9584.
Recent searches for water masers in six lensed quasars discovered a maser in MG 0414+0534 \citep{maser,mckean}, indicating the most distant water found in the universe. 
The emission comes from the $6_{16}-5_{23}$ transition of H$_2$O at a rest frequency of 22.235 GHz.
 The observations show a line with a single   peak, Doppler broadened by about  $\sim$ 100 km s$^{-1}$, indicating a spatially extended maser source, offset from the systemic velocity of the quasar by about -300 km s$^{-1}$. There are four images of the quasar produced in the MG 0414+0534  system,
designated A1, A2, B and C, and the maser is observed in A1 and A2. \hl{ The sensitivity of the telescope was not enough
to detect the maser emission in images B and C, which are known to be fainter than A1 and A2.}

Most known
galactic masers are believed to fall into two categories: disk  or jet masers, both of which may be found 
in the same galaxy and   usually consist of groups of ``spots''. Disk masers are generated in a rotating molecular disk
 around the central engine of an active galaxy and produce a characteristic three-peaked spectrum with blue- and red-shifted components. The canonical
example of this configuration is NGC 4258 \citep{moran}. 
Jet masers  are probably generated in turbulent
shock fronts around the edges,  or at 
the ends, of  AGN radio jets.  Such a group of masers will be located some way along the jet and be wholly red- or blue-shifted depending on the jet direction. The masers
are typically offset from the AGN core by $\sim$ 100 km s$^{-1}$,
and  have a radial-velocity gradient of $\sim$ 100 km s$^{-1}$mas$^{-1}$  in a single direction across the group.
\shl The velocity of the masers is not likely to be strongly correlated with the jet velocity for a couple of reasons. 
 \shl Firstly, \shl the 
jet speed
is likely to be \shlA a significant \ehlA fraction of the speed of light \citep{ros}, and masers could not survive in such an environment; 
\shl secondly,  \shl the direction of the masers may be influenced 
by  turbulent motion near the jet. \ehl Examples of 
galaxies with jet masers are
NGC 1052 \citep{ngc1052} and Mrk 348 \citep{mrk348}.
Based on these two categories, the maser in MG 0414+0534 is most likely to be
a jet maser group, since there is only one peak, blue-shifted from the core by 300 km s$^{-1}$, and a velocity spread of $\sim$100 km s$^{-1}$. It could be 
in a jet that is inclined towards us, which is consistent with MG 0414+0534 being a type-1 quasar \citep{lawrence}. 

\subsection{Analysis of microlensed quasars}
\label{analysis_described}
We will adopt a method that has been used previously to study  microlensing in MG 0414+0534 and other lensed quasars \citep[e.g.][]{wambsganss_2237,schechter1,schechter2,lewis1,bate1}. Firstly we  calculate the microlensing of a point-like 
source (i.e. the smallest numerically possible) situated
in the \emph{source plane} behind the lensing galaxy. The mass distribution in the lens, producing each image, is specified by a convergence ($\kappa$) and a shear ($\gamma$). The convergence  describes the effect of 
mass near a light ray, and the shear incorporates long-range influences of the overall  mass distribution.
Values for these parameters for many quasars, including MG 0414+0534, have been determined previously \citep{witt,schechter1}.
Based on these, point masses are  laid down randomly over another grid that is the \emph{lens plane}.  \emph{Ray-tracing} is then used to find the path of the light as it
travels from the source, is deflected through the lens, and reaches the observer. In practice, however, it is more computationally efficient  to use light rays that are fired back from
the observer through the lens to the source plane. This \emph{inverse ray-tracing} method simultaneously calculates the magnification of the
point source for many locations over the source plane.  

The result of
inverse ray-tracing is a \emph{magnification map} of the source plane, as shown in Figure \ref{big_map}. One map is needed for each image. A map does not encode an image of the quasar that would be seen by  us. Rather, each point on the map represents the
magnification of a pixel-sized source placed at that location, for all locations over the source plane region being modelled. Brighter areas are high magnification regions and darker areas are low magnification regions, and are relative to a mean magnification value.  Notice that in the map there are thick lines or ``trunks'' of high magnification running up and down, these are the effects of the shear  which imposes a directional trend on the map. It is not known what the true shear direction is for MG 0414+0534. The break-out box shows a region of the map at high
resolution; one can see intricate patterns of
light and dark which are called \emph{caustics}. As a source moves across these the magnification varies, particularly
when the source crosses a bright edge, leading to high magnification events. \shl The  Einstein Radius is defined in the next subsection.\ehl

Spatially extended  sources are studied by convolving source models with the magnification map. To study the \emph{differential microlensing} \citep{keeton} of frequencies emitted from separate regions of an extended source,  each region is cut from the source and convolved separately to obtain a magnification for that
frequency, and then combined to produce a microlensed spectrum.
If this is done for points along a path in the source plane then a time-varying
spectrum for a source is obtained.

\subsection{Microlensing distance and time scales} \shl

Microlensing systems are characterised by several time and spatial scales: the time between high magnification events,  the duration of 
their occurrence, and  the characteristic  distance scale: the \emph{Einstein Radius}.
An Einstein Ring is generated  from the simplest lensing situation,  where a point source is directly behind a  point
lens, both perfectly in line with the observer. In that case the image that is produced is a  circle  around the lens. The 
observed angle $\theta_E$ between the lens and the circle circumference  is given by 
\begin{equation}
\label{er_eqn}
\theta_E = \sqrt{ \frac{4 G M }{c^2} \frac{D_{LS}}{D_L D_S} },
\end{equation}
where  $c$ the speed of light, $G$ the gravitational constant, $M$ the lens mass,
$D_{LS}, D_L$ and  $D_S$ the distance from lens to source, observer to lens, and observer to source, respectively. 
Projecting $\theta_E$  onto the source plane
produces a distance called the Einstein Radius  (ER). Thus,
the ER incorporates many aspects of the lens system, and is \shl used  as a length scale in microlensing analysis. 
 It also indicates the size scale \shlA relevant \ehlA for sources; sources of  order 1 ER in size, and smaller,  are more likely to be
susceptible to microlensing \shlA than larger sources \ehlA (see \citet{refsdal1,refsdal2} for the microlensing of large sources). For  MG 0414+0534   we calculate the ER based on a point mass of 1 $M_\odot$, since that size object will be used in our model galaxy. With a redshift for the lens of z = 0.9584 and
for the quasar of z = 2.639, and assuming a concordance cosmology with ${\Omega}_0 = 0.3$, ${\Lambda}_0 = 0.7$ and $H_0 = 72$ 
km s$^{-1}$ Mpc$^{-1}$, then 1 ER  $\simeq$ \shlA 0.013 pc. \ehlA 
The ER can be converted to a time scale if the relative velocity of the observer, lens, and source is known. Changes over time, rather than the ER,
represent an observable of the lens system.
\citep{chartas} derive a velocity of 170 km s$^{-1}$ for converting source distance to time in MG0414+0534, so the time to travel 1 ER is 77 years. 
\ehl

\section{Method}
\label{Method}
We  use the inverse ray-shooting method \citep{Kayser}  developed by  \cite{wambsganss_comp} and
parallelized by \cite{garsden}, to produce the magnification maps. Source models (detailed below)
are convolved using Fourier transforms. All of these
steps are executed using parallel processes and distributed memory on a supercomputer. Spectra are then extracted from a line across the
map chosen to exhibit variability, and plotted for graphical examination. Parameters are extracted and reduced to measures of variability.

\subsection{Magnification Maps}
The convergence and shear  for MG 0414+0534 are taken  from \citet{witt} and  listed in Table \ref{parameters}.
The maser has been observed in images A1 and A2, so only these will be modelled. For each image, two magnification
maps are used with different resolutions, because the models represent masers  of various sizes.
The highest resolution maps span a source plane region of 
$2.5 \times 2.5$ pc$^2$ ($185 \times 185$ ER$^2$), the  lower resolution maps cover  $15 \times 15$ pc$^2$ 
($1115 \times 1115$ ER$^2$). All  maps  contain $20000 \times 20000$ pixels giving a source size resolution   of  $26\times26$ AU$^2$ and $155\times 155$ AU$^2$ respectively, around the size of the solar system. All objects comprising the lens have masses of 1 solar mass (M$_\odot$). The number of objects is determined
by the lens model and reaches 241,562,900 for the low resolution  map for image A2. The map shown in Figure \ref{big_map}
 is the $2.5 \times 2.5$ pc$^2$ map for image A1, and the A2 map is similar in appearance.
We assign directions to the map so that the orientation of maser models can be discussed. North (N) points up and East (E) is to the right, as in a land map. The directions are significant
because they relate to the direction of shear that produces vertical patterns in the maps, which by our convention run N-S. 

\subsection{Maser source models}

To model lensing of the maser in MG 0414+0534 we have chosen several types of source models of varying levels of complexity. We assume the maser material is transparent, and emitting
isotropically. \hl{Because we wish to study the effect of microlensing on the spectrum and total flux, each model needs a size, a spatial flux emission profile, and a spatial frequency emission profile. The first two are  described
in Figure \ref{schemes}.
The third is specified by the velocity profile of the source components, which translates to a frequency emission profile due to Doppler shifts. } However, the maser velocities are not well-understood, as discussed in Section \ref{lens and maser}.
Therefore we use the geometry of the model, and estimates based on observations, to obtain velocities for the
masers, but scale them so  that the line-of-sight velocity falls in a range of 100 km s$^{-1}$ (except for the Single Spot which has no range). This  is then sliced into 40 velocity intervals to produce a spectrum. \hl{ All the model descriptions and results will refer to the velocities in the model, rather than frequencies.}
 The next sections describe the different models.

\subsubsection{Single Spot}
This is the simplest model and is a single maser spot implemented as a 2-D Gaussian of total width 0.002 pc with a central velocity of -300 km s$^{-1}$ and no velocity broadening. Since MG 0414+0534  is  a powerful quasar we may assume that the spot size is at
the high end of expected sizes for jet masers, so we chose the value of 0.002 pc, which also may 
 approximate that of the spots  in NGC 1052 \citep{ngc1052}.

\begin{figure*}
\centering    
\begingroup%
\makeatletter%
\newcommand{\GNUPLOTspecial}{%
  \@sanitize\catcode`\%=14\relax\special}%
\setlength{\unitlength}{0.0500bp}%
\begin{picture}(5760,4320)(0,0)%
  {\GNUPLOTspecial{"
/gnudict 256 dict def
gnudict begin
%
%
/Color false def
/Blacktext true def
/Solid false def
/Dashlength 1 def
/Landscape false def
/Level1 false def
/Rounded false def
/ClipToBoundingBox false def
/TransparentPatterns false def
/gnulinewidth 7.000 def
/userlinewidth gnulinewidth def
/Gamma 1.0 def
/vshift -66 def
/dl1 {
  10.0 Dashlength mul mul
  Rounded { currentlinewidth 0.75 mul sub dup 0 le { pop 0.01 } if } if
} def
/dl2 {
  10.0 Dashlength mul mul
  Rounded { currentlinewidth 0.75 mul add } if
} def
/hpt_ 31.5 def
/vpt_ 31.5 def
/hpt hpt_ def
/vpt vpt_ def
Level1 {} {
/SDict 10 dict def
systemdict /pdfmark known not {
  userdict /pdfmark systemdict /cleartomark get put
} if
SDict begin [
  /Title (histogram.tex)
  /Subject (gnuplot plot)
  /Creator (gnuplot 4.4 patchlevel 0)
  /Author (hgarsden)
  /CreationDate (Wed Oct  6 19:01:05 2010)
  /DOCINFO pdfmark
end
} ifelse
/doclip {
  ClipToBoundingBox {
    newpath 0 0 moveto 288 0 lineto 288 216 lineto 0 216 lineto closepath
    clip
  } if
} def
%
%
%
/M {moveto} bind def
/L {lineto} bind def
/R {rmoveto} bind def
/V {rlineto} bind def
/N {newpath moveto} bind def
/Z {closepath} bind def
/C {setrgbcolor} bind def
/f {rlineto fill} bind def
/Gshow {show} def   
/vpt2 vpt 2 mul def
/hpt2 hpt 2 mul def
/Lshow {currentpoint stroke M 0 vshift R 
	Blacktext {gsave 0 setgray show grestore} {show} ifelse} def
/Rshow {currentpoint stroke M dup stringwidth pop neg vshift R
	Blacktext {gsave 0 setgray show grestore} {show} ifelse} def
/Cshow {currentpoint stroke M dup stringwidth pop -2 div vshift R 
	Blacktext {gsave 0 setgray show grestore} {show} ifelse} def
/UP {dup vpt_ mul /vpt exch def hpt_ mul /hpt exch def
  /hpt2 hpt 2 mul def /vpt2 vpt 2 mul def} def
/DL {Color {setrgbcolor Solid {pop []} if 0 setdash}
 {pop pop pop 0 setgray Solid {pop []} if 0 setdash} ifelse} def
/BL {stroke userlinewidth 2 mul setlinewidth
	Rounded {1 setlinejoin 1 setlinecap} if} def
/AL {stroke userlinewidth 2 div setlinewidth
	Rounded {1 setlinejoin 1 setlinecap} if} def
/UL {dup gnulinewidth mul /userlinewidth exch def
	dup 1 lt {pop 1} if 10 mul /udl exch def} def
/PL {stroke userlinewidth setlinewidth
	Rounded {1 setlinejoin 1 setlinecap} if} def
/LCw {1 1 1} def
/LCb {0 0 0} def
/LCa {0 0 0} def
/LC0 {1 0 0} def
/LC1 {0 1 0} def
/LC2 {0 0 1} def
/LC3 {1 0 1} def
/LC4 {0 1 1} def
/LC5 {1 1 0} def
/LC6 {0 0 0} def
/LC7 {1 0.3 0} def
/LC8 {0.5 0.5 0.5} def
/LTw {PL [] 1 setgray} def
/LTb {BL [] LCb DL} def
/LTa {AL [1 udl mul 2 udl mul] 0 setdash LCa setrgbcolor} def
/LT0 {PL [] LC0 DL} def
/LT1 {PL [4 dl1 2 dl2] LC1 DL} def
/LT2 {PL [2 dl1 3 dl2] LC2 DL} def
/LT3 {PL [1 dl1 1.5 dl2] LC3 DL} def
/LT4 {PL [6 dl1 2 dl2 1 dl1 2 dl2] LC4 DL} def
/LT5 {PL [3 dl1 3 dl2 1 dl1 3 dl2] LC5 DL} def
/LT6 {PL [2 dl1 2 dl2 2 dl1 6 dl2] LC6 DL} def
/LT7 {PL [1 dl1 2 dl2 6 dl1 2 dl2 1 dl1 2 dl2] LC7 DL} def
/LT8 {PL [2 dl1 2 dl2 2 dl1 2 dl2 2 dl1 2 dl2 2 dl1 4 dl2] LC8 DL} def
/Pnt {stroke [] 0 setdash gsave 1 setlinecap M 0 0 V stroke grestore} def
/Dia {stroke [] 0 setdash 2 copy vpt add M
  hpt neg vpt neg V hpt vpt neg V
  hpt vpt V hpt neg vpt V closepath stroke
  Pnt} def
/Pls {stroke [] 0 setdash vpt sub M 0 vpt2 V
  currentpoint stroke M
  hpt neg vpt neg R hpt2 0 V stroke
 } def
/Box {stroke [] 0 setdash 2 copy exch hpt sub exch vpt add M
  0 vpt2 neg V hpt2 0 V 0 vpt2 V
  hpt2 neg 0 V closepath stroke
  Pnt} def
/Crs {stroke [] 0 setdash exch hpt sub exch vpt add M
  hpt2 vpt2 neg V currentpoint stroke M
  hpt2 neg 0 R hpt2 vpt2 V stroke} def
/TriU {stroke [] 0 setdash 2 copy vpt 1.12 mul add M
  hpt neg vpt -1.62 mul V
  hpt 2 mul 0 V
  hpt neg vpt 1.62 mul V closepath stroke
  Pnt} def
/Star {2 copy Pls Crs} def
/BoxF {stroke [] 0 setdash exch hpt sub exch vpt add M
  0 vpt2 neg V hpt2 0 V 0 vpt2 V
  hpt2 neg 0 V closepath fill} def
/TriUF {stroke [] 0 setdash vpt 1.12 mul add M
  hpt neg vpt -1.62 mul V
  hpt 2 mul 0 V
  hpt neg vpt 1.62 mul V closepath fill} def
/TriD {stroke [] 0 setdash 2 copy vpt 1.12 mul sub M
  hpt neg vpt 1.62 mul V
  hpt 2 mul 0 V
  hpt neg vpt -1.62 mul V closepath stroke
  Pnt} def
/TriDF {stroke [] 0 setdash vpt 1.12 mul sub M
  hpt neg vpt 1.62 mul V
  hpt 2 mul 0 V
  hpt neg vpt -1.62 mul V closepath fill} def
/DiaF {stroke [] 0 setdash vpt add M
  hpt neg vpt neg V hpt vpt neg V
  hpt vpt V hpt neg vpt V closepath fill} def
/Pent {stroke [] 0 setdash 2 copy gsave
  translate 0 hpt M 4 {72 rotate 0 hpt L} repeat
  closepath stroke grestore Pnt} def
/PentF {stroke [] 0 setdash gsave
  translate 0 hpt M 4 {72 rotate 0 hpt L} repeat
  closepath fill grestore} def
/Circle {stroke [] 0 setdash 2 copy
  hpt 0 360 arc stroke Pnt} def
/CircleF {stroke [] 0 setdash hpt 0 360 arc fill} def
/C0 {BL [] 0 setdash 2 copy moveto vpt 90 450 arc} bind def
/C1 {BL [] 0 setdash 2 copy moveto
	2 copy vpt 0 90 arc closepath fill
	vpt 0 360 arc closepath} bind def
/C2 {BL [] 0 setdash 2 copy moveto
	2 copy vpt 90 180 arc closepath fill
	vpt 0 360 arc closepath} bind def
/C3 {BL [] 0 setdash 2 copy moveto
	2 copy vpt 0 180 arc closepath fill
	vpt 0 360 arc closepath} bind def
/C4 {BL [] 0 setdash 2 copy moveto
	2 copy vpt 180 270 arc closepath fill
	vpt 0 360 arc closepath} bind def
/C5 {BL [] 0 setdash 2 copy moveto
	2 copy vpt 0 90 arc
	2 copy moveto
	2 copy vpt 180 270 arc closepath fill
	vpt 0 360 arc} bind def
/C6 {BL [] 0 setdash 2 copy moveto
	2 copy vpt 90 270 arc closepath fill
	vpt 0 360 arc closepath} bind def
/C7 {BL [] 0 setdash 2 copy moveto
	2 copy vpt 0 270 arc closepath fill
	vpt 0 360 arc closepath} bind def
/C8 {BL [] 0 setdash 2 copy moveto
	2 copy vpt 270 360 arc closepath fill
	vpt 0 360 arc closepath} bind def
/C9 {BL [] 0 setdash 2 copy moveto
	2 copy vpt 270 450 arc closepath fill
	vpt 0 360 arc closepath} bind def
/C10 {BL [] 0 setdash 2 copy 2 copy moveto vpt 270 360 arc closepath fill
	2 copy moveto
	2 copy vpt 90 180 arc closepath fill
	vpt 0 360 arc closepath} bind def
/C11 {BL [] 0 setdash 2 copy moveto
	2 copy vpt 0 180 arc closepath fill
	2 copy moveto
	2 copy vpt 270 360 arc closepath fill
	vpt 0 360 arc closepath} bind def
/C12 {BL [] 0 setdash 2 copy moveto
	2 copy vpt 180 360 arc closepath fill
	vpt 0 360 arc closepath} bind def
/C13 {BL [] 0 setdash 2 copy moveto
	2 copy vpt 0 90 arc closepath fill
	2 copy moveto
	2 copy vpt 180 360 arc closepath fill
	vpt 0 360 arc closepath} bind def
/C14 {BL [] 0 setdash 2 copy moveto
	2 copy vpt 90 360 arc closepath fill
	vpt 0 360 arc} bind def
/C15 {BL [] 0 setdash 2 copy vpt 0 360 arc closepath fill
	vpt 0 360 arc closepath} bind def
/Rec {newpath 4 2 roll moveto 1 index 0 rlineto 0 exch rlineto
	neg 0 rlineto closepath} bind def
/Square {dup Rec} bind def
/Bsquare {vpt sub exch vpt sub exch vpt2 Square} bind def
/S0 {BL [] 0 setdash 2 copy moveto 0 vpt rlineto BL Bsquare} bind def
/S1 {BL [] 0 setdash 2 copy vpt Square fill Bsquare} bind def
/S2 {BL [] 0 setdash 2 copy exch vpt sub exch vpt Square fill Bsquare} bind def
/S3 {BL [] 0 setdash 2 copy exch vpt sub exch vpt2 vpt Rec fill Bsquare} bind def
/S4 {BL [] 0 setdash 2 copy exch vpt sub exch vpt sub vpt Square fill Bsquare} bind def
/S5 {BL [] 0 setdash 2 copy 2 copy vpt Square fill
	exch vpt sub exch vpt sub vpt Square fill Bsquare} bind def
/S6 {BL [] 0 setdash 2 copy exch vpt sub exch vpt sub vpt vpt2 Rec fill Bsquare} bind def
/S7 {BL [] 0 setdash 2 copy exch vpt sub exch vpt sub vpt vpt2 Rec fill
	2 copy vpt Square fill Bsquare} bind def
/S8 {BL [] 0 setdash 2 copy vpt sub vpt Square fill Bsquare} bind def
/S9 {BL [] 0 setdash 2 copy vpt sub vpt vpt2 Rec fill Bsquare} bind def
/S10 {BL [] 0 setdash 2 copy vpt sub vpt Square fill 2 copy exch vpt sub exch vpt Square fill
	Bsquare} bind def
/S11 {BL [] 0 setdash 2 copy vpt sub vpt Square fill 2 copy exch vpt sub exch vpt2 vpt Rec fill
	Bsquare} bind def
/S12 {BL [] 0 setdash 2 copy exch vpt sub exch vpt sub vpt2 vpt Rec fill Bsquare} bind def
/S13 {BL [] 0 setdash 2 copy exch vpt sub exch vpt sub vpt2 vpt Rec fill
	2 copy vpt Square fill Bsquare} bind def
/S14 {BL [] 0 setdash 2 copy exch vpt sub exch vpt sub vpt2 vpt Rec fill
	2 copy exch vpt sub exch vpt Square fill Bsquare} bind def
/S15 {BL [] 0 setdash 2 copy Bsquare fill Bsquare} bind def
/D0 {gsave translate 45 rotate 0 0 S0 stroke grestore} bind def
/D1 {gsave translate 45 rotate 0 0 S1 stroke grestore} bind def
/D2 {gsave translate 45 rotate 0 0 S2 stroke grestore} bind def
/D3 {gsave translate 45 rotate 0 0 S3 stroke grestore} bind def
/D4 {gsave translate 45 rotate 0 0 S4 stroke grestore} bind def
/D5 {gsave translate 45 rotate 0 0 S5 stroke grestore} bind def
/D6 {gsave translate 45 rotate 0 0 S6 stroke grestore} bind def
/D7 {gsave translate 45 rotate 0 0 S7 stroke grestore} bind def
/D8 {gsave translate 45 rotate 0 0 S8 stroke grestore} bind def
/D9 {gsave translate 45 rotate 0 0 S9 stroke grestore} bind def
/D10 {gsave translate 45 rotate 0 0 S10 stroke grestore} bind def
/D11 {gsave translate 45 rotate 0 0 S11 stroke grestore} bind def
/D12 {gsave translate 45 rotate 0 0 S12 stroke grestore} bind def
/D13 {gsave translate 45 rotate 0 0 S13 stroke grestore} bind def
/D14 {gsave translate 45 rotate 0 0 S14 stroke grestore} bind def
/D15 {gsave translate 45 rotate 0 0 S15 stroke grestore} bind def
/DiaE {stroke [] 0 setdash vpt add M
  hpt neg vpt neg V hpt vpt neg V
  hpt vpt V hpt neg vpt V closepath stroke} def
/BoxE {stroke [] 0 setdash exch hpt sub exch vpt add M
  0 vpt2 neg V hpt2 0 V 0 vpt2 V
  hpt2 neg 0 V closepath stroke} def
/TriUE {stroke [] 0 setdash vpt 1.12 mul add M
  hpt neg vpt -1.62 mul V
  hpt 2 mul 0 V
  hpt neg vpt 1.62 mul V closepath stroke} def
/TriDE {stroke [] 0 setdash vpt 1.12 mul sub M
  hpt neg vpt 1.62 mul V
  hpt 2 mul 0 V
  hpt neg vpt -1.62 mul V closepath stroke} def
/PentE {stroke [] 0 setdash gsave
  translate 0 hpt M 4 {72 rotate 0 hpt L} repeat
  closepath stroke grestore} def
/CircE {stroke [] 0 setdash 
  hpt 0 360 arc stroke} def
/Opaque {gsave closepath 1 setgray fill grestore 0 setgray closepath} def
/DiaW {stroke [] 0 setdash vpt add M
  hpt neg vpt neg V hpt vpt neg V
  hpt vpt V hpt neg vpt V Opaque stroke} def
/BoxW {stroke [] 0 setdash exch hpt sub exch vpt add M
  0 vpt2 neg V hpt2 0 V 0 vpt2 V
  hpt2 neg 0 V Opaque stroke} def
/TriUW {stroke [] 0 setdash vpt 1.12 mul add M
  hpt neg vpt -1.62 mul V
  hpt 2 mul 0 V
  hpt neg vpt 1.62 mul V Opaque stroke} def
/TriDW {stroke [] 0 setdash vpt 1.12 mul sub M
  hpt neg vpt 1.62 mul V
  hpt 2 mul 0 V
  hpt neg vpt -1.62 mul V Opaque stroke} def
/PentW {stroke [] 0 setdash gsave
  translate 0 hpt M 4 {72 rotate 0 hpt L} repeat
  Opaque stroke grestore} def
/CircW {stroke [] 0 setdash 
  hpt 0 360 arc Opaque stroke} def
/BoxFill {gsave Rec 1 setgray fill grestore} def
/Density {
  /Fillden exch def
  currentrgbcolor
  /ColB exch def /ColG exch def /ColR exch def
  /ColR ColR Fillden mul Fillden sub 1 add def
  /ColG ColG Fillden mul Fillden sub 1 add def
  /ColB ColB Fillden mul Fillden sub 1 add def
  ColR ColG ColB setrgbcolor} def
/BoxColFill {gsave Rec PolyFill} def
/PolyFill {gsave Density fill grestore grestore} def
/h {rlineto rlineto rlineto gsave closepath fill grestore} bind def
%
%
/PatternFill {gsave /PFa [ 9 2 roll ] def
  PFa 0 get PFa 2 get 2 div add PFa 1 get PFa 3 get 2 div add translate
  PFa 2 get -2 div PFa 3 get -2 div PFa 2 get PFa 3 get Rec
  gsave 1 setgray fill grestore clip
  currentlinewidth 0.5 mul setlinewidth
  /PFs PFa 2 get dup mul PFa 3 get dup mul add sqrt def
  0 0 M PFa 5 get rotate PFs -2 div dup translate
  0 1 PFs PFa 4 get div 1 add floor cvi
	{PFa 4 get mul 0 M 0 PFs V} for
  0 PFa 6 get ne {
	0 1 PFs PFa 4 get div 1 add floor cvi
	{PFa 4 get mul 0 2 1 roll M PFs 0 V} for
 } if
  stroke grestore} def
/languagelevel where
 {pop languagelevel} {1} ifelse
 2 lt
	{/InterpretLevel1 true def}
	{/InterpretLevel1 Level1 def}
 ifelse
%
%
/Level2PatternFill {
/Tile8x8 {/PaintType 2 /PatternType 1 /TilingType 1 /BBox [0 0 8 8] /XStep 8 /YStep 8}
	bind def
/KeepColor {currentrgbcolor [/Pattern /DeviceRGB] setcolorspace} bind def
<< Tile8x8
 /PaintProc {0.5 setlinewidth pop 0 0 M 8 8 L 0 8 M 8 0 L stroke} 
>> matrix makepattern
/Pat1 exch def
<< Tile8x8
 /PaintProc {0.5 setlinewidth pop 0 0 M 8 8 L 0 8 M 8 0 L stroke
	0 4 M 4 8 L 8 4 L 4 0 L 0 4 L stroke}
>> matrix makepattern
/Pat2 exch def
<< Tile8x8
 /PaintProc {0.5 setlinewidth pop 0 0 M 0 8 L
	8 8 L 8 0 L 0 0 L fill}
>> matrix makepattern
/Pat3 exch def
<< Tile8x8
 /PaintProc {0.5 setlinewidth pop -4 8 M 8 -4 L
	0 12 M 12 0 L stroke}
>> matrix makepattern
/Pat4 exch def
<< Tile8x8
 /PaintProc {0.5 setlinewidth pop -4 0 M 8 12 L
	0 -4 M 12 8 L stroke}
>> matrix makepattern
/Pat5 exch def
<< Tile8x8
 /PaintProc {0.5 setlinewidth pop -2 8 M 4 -4 L
	0 12 M 8 -4 L 4 12 M 10 0 L stroke}
>> matrix makepattern
/Pat6 exch def
<< Tile8x8
 /PaintProc {0.5 setlinewidth pop -2 0 M 4 12 L
	0 -4 M 8 12 L 4 -4 M 10 8 L stroke}
>> matrix makepattern
/Pat7 exch def
<< Tile8x8
 /PaintProc {0.5 setlinewidth pop 8 -2 M -4 4 L
	12 0 M -4 8 L 12 4 M 0 10 L stroke}
>> matrix makepattern
/Pat8 exch def
<< Tile8x8
 /PaintProc {0.5 setlinewidth pop 0 -2 M 12 4 L
	-4 0 M 12 8 L -4 4 M 8 10 L stroke}
>> matrix makepattern
/Pat9 exch def
/Pattern1 {PatternBgnd KeepColor Pat1 setpattern} bind def
/Pattern2 {PatternBgnd KeepColor Pat2 setpattern} bind def
/Pattern3 {PatternBgnd KeepColor Pat3 setpattern} bind def
/Pattern4 {PatternBgnd KeepColor Landscape {Pat5} {Pat4} ifelse setpattern} bind def
/Pattern5 {PatternBgnd KeepColor Landscape {Pat4} {Pat5} ifelse setpattern} bind def
/Pattern6 {PatternBgnd KeepColor Landscape {Pat9} {Pat6} ifelse setpattern} bind def
/Pattern7 {PatternBgnd KeepColor Landscape {Pat8} {Pat7} ifelse setpattern} bind def
} def
%
%
%
/PatternBgnd {
  TransparentPatterns {} {gsave 1 setgray fill grestore} ifelse
} def
%
%
/Level1PatternFill {
/Pattern1 {0.250 Density} bind def
/Pattern2 {0.500 Density} bind def
/Pattern3 {0.750 Density} bind def
/Pattern4 {0.125 Density} bind def
/Pattern5 {0.375 Density} bind def
/Pattern6 {0.625 Density} bind def
/Pattern7 {0.875 Density} bind def
} def
%
%
Level1 {Level1PatternFill} {Level2PatternFill} ifelse
/Symbol-Oblique /Symbol findfont [1 0 .167 1 0 0] makefont
dup length dict begin {1 index /FID eq {pop pop} {def} ifelse} forall
currentdict end definefont pop
end
gnudict begin
gsave
doclip
0 0 translate
0.050 0.050 scale
0 setgray
newpath
1.000 UL
LTb
0 2160 M
0 37 V
0 4319 M
0 -37 V
480 2160 M
0 37 V
0 2122 R
0 -37 V
960 2160 M
0 37 V
0 2122 R
0 -37 V
1440 2160 M
0 37 V
0 2122 R
0 -37 V
1919 2160 M
0 37 V
0 2122 R
0 -37 V
2399 2160 M
0 37 V
0 2122 R
0 -37 V
2879 2160 M
0 37 V
0 2122 R
0 -37 V
stroke
0 4319 N
0 2160 L
2879 0 V
0 2159 V
0 4319 L
Z stroke
LCb setrgbcolor
LTb
1.000 UP
1.000 UL
LTb
LCb setrgbcolor
LTb
1.000 UL
LTb
60 2160 M
60 0 V
60 0 V
60 0 V
60 0 V
60 0 V
60 0 V
60 0 V
60 0 V
60 0 V
60 0 V
60 0 V
60 0 V
60 0 V
60 0 V
60 0 V
60 0 V
60 0 V
60 0 V
60 0 V
60 0 V
60 14 V
60 1771 V
60 181 V
59 -1944 V
60 -22 V
60 0 V
60 0 V
60 0 V
60 0 V
60 0 V
60 0 V
60 0 V
60 0 V
60 0 V
60 0 V
60 0 V
60 0 V
60 0 V
60 0 V
60 0 V
60 0 V
60 0 V
60 0 V
60 0 V
60 0 V
stroke
0 4319 N
0 2160 L
2879 0 V
0 2159 V
0 4319 L
Z stroke
1.000 UP
1.000 UL
LTb
1.000 UL
LTb
2880 2160 M
0 37 V
0 2122 R
0 -37 V
3360 2160 M
0 37 V
0 2122 R
0 -37 V
3840 2160 M
0 37 V
0 2122 R
0 -37 V
4320 2160 M
0 37 V
0 2122 R
0 -37 V
4799 2160 M
0 37 V
0 2122 R
0 -37 V
5279 2160 M
0 37 V
0 2122 R
0 -37 V
5759 2160 M
0 37 V
0 2122 R
0 -37 V
-2879 37 R
0 -2159 R
2879 0 R
0 2159 V
-2879 0 V
1.000 UP
stroke
LCb setrgbcolor
LTb
1.000 UL
LTb
2940 2160 M
60 0 V
60 0 V
48 5 V
12 171 V
60 50 V
60 60 V
60 68 V
60 80 V
60 91 V
60 101 V
60 112 V
60 121 V
60 129 V
60 133 V
60 135 V
60 134 V
60 129 V
60 120 V
60 107 V
60 91 V
60 71 V
60 49 V
60 25 V
60 0 V
59 -24 V
60 -49 V
60 -71 V
60 -90 V
60 -107 V
60 -120 V
60 -128 V
60 -134 V
60 -135 V
60 -134 V
60 -128 V
60 -121 V
60 -112 V
60 -102 V
60 -91 V
60 -80 V
60 -69 V
60 -60 V
60 -50 V
24 -172 V
96 -5 V
60 0 V
2880 4319 M
0 -2159 R
2879 0 R
0 2159 V
-2879 0 V
1.000 UP
0 0 M
0 37 L
0 2160 M
0 -37 V
480 0 M
0 37 V
0 2123 R
0 -37 V
960 0 M
0 37 V
0 2123 R
0 -37 V
1440 0 M
0 37 V
0 2123 R
0 -37 V
1919 0 M
0 37 V
0 2123 R
0 -37 V
2399 0 M
0 37 V
0 2123 R
0 -37 V
2879 0 M
0 37 V
0 2123 R
0 -37 V
0 2160 M
0 0 L
2879 0 L
0 2160 V
0 2160 M
1.000 UP
stroke
LCb setrgbcolor
LTb
1.000 UL
LTb
60 0 M
60 0 V
60 0 V
60 0 V
60 0 V
60 0 V
60 4 V
60 237 V
60 -51 V
60 208 V
60 30 V
720 88 L
60 240 V
840 165 L
60 347 V
60 356 V
60 -192 V
60 84 V
60 -326 V
60 -120 V
60 790 V
60 -816 V
60 -164 V
60 273 V
59 -40 V
60 154 V
60 -43 V
60 -121 V
60 809 V
60 551 V
60 255 V
1919 888 L
60 -342 V
60 -343 V
60 316 V
60 -10 V
60 -161 V
60 -115 V
60 431 V
60 -551 V
12 -60 V
12 0 V
12 -32 V
12 0 V
2459 0 L
60 0 V
60 0 V
60 0 V
60 0 V
60 0 V
0 2160 M
0 0 L
2879 0 L
0 2160 V
0 2160 M
1.000 UP
2880 0 M
0 37 V
0 2123 R
0 -37 V
3360 0 M
0 37 V
0 2123 R
0 -37 V
3840 0 M
0 37 V
0 2123 R
0 -37 V
4320 0 M
0 37 V
0 2123 R
0 -37 V
4799 0 M
0 37 V
0 2123 R
0 -37 V
5279 0 M
0 37 V
0 2123 R
0 -37 V
5759 0 M
0 37 V
0 2123 R
0 -37 V
-2879 37 R
2880 0 M
5759 0 L
0 2160 V
-2879 0 V
stroke
LCb setrgbcolor
LTb
1.000 UP
1.000 UL
LTb
LCb setrgbcolor
LTb
1.000 UL
LTb
2940 0 M
60 0 V
60 0 V
60 14 V
60 23 V
60 751 V
60 808 V
60 -507 V
60 -424 V
60 -28 V
60 -30 V
60 21 V
60 -262 V
60 198 V
60 -57 V
60 -169 V
60 229 V
60 -234 V
60 102 V
60 -57 V
60 55 V
60 -56 V
60 14 V
60 -132 V
59 109 V
60 60 V
60 -1 V
60 -62 V
60 81 V
60 -14 V
60 -74 V
60 47 V
60 146 V
60 -6 V
60 -61 V
60 28 V
60 -65 V
60 392 V
60 -132 V
60 274 V
60 961 V
5399 419 L
60 -282 V
5519 14 L
5579 0 L
60 0 V
2880 2160 M
2880 0 M
5759 0 L
0 2160 V
-2879 0 V
1.000 UP
stroke
grestore
end
showpage
  }}%
  \put(3891,1997){\makebox(0,0){\strut{}Grainy Ring $i=50\,^\circ$ East}}%
  \put(2919,-499){\makebox(0,0){\strut{}Velocity}}%
  \put(5279,-200){\makebox(0,0){\strut{}-260}}%
  \put(4799,-200){\makebox(0,0){\strut{}-280}}%
  \put(4320,-200){\makebox(0,0){\strut{}-300}}%
  \put(3840,-200){\makebox(0,0){\strut{}-320}}%
  \put(3360,-200){\makebox(0,0){\strut{}-340}}%
  \put(531,1997){\makebox(0,0){\strut{}Grainy Patch}}%
  \put(2399,-200){\makebox(0,0){\strut{}-260}}%
  \put(1919,-200){\makebox(0,0){\strut{}-280}}%
  \put(1440,-200){\makebox(0,0){\strut{}-300}}%
  \put(960,-200){\makebox(0,0){\strut{}-320}}%
  \put(480,-200){\makebox(0,0){\strut{}-340}}%
  \put(3451,4156){\makebox(0,0){\strut{}Smooth Patch}}%
  \put(471,4156){\makebox(0,0){\strut{}Single Spot}}%
  \put(-200,2239){%
  \makebox(0,0){\strut{}Flux}%
  }%
\end{picture}%
\endgroup
\vspace{10mm}
\caption{The spectra of the four simplest source models without lensing. The y-axis
is the flux, the x-axis is the velocity present in the source model. Reading left to right, top to bottom,  the spectra are for the models: Single Spot, Smooth Patch, 
Grainy Patch, Grainy Ring $i=50\,^{\circ}$ East. The Single Spot has no velocity spread because we have not  assigned it any, and produces a sharp peak (within the resolution of the slices). The Smooth Patch is a Gaussian and produces
a Gaussian spectrum. The patch of spots, each one a Single Spot, produces a spectrum of many peaks reflecting the distribution of spots; at the right there is a region where there are a few more spots and which produces the highest peak. The ring
produces a typical ring-generated spectrum with peaks at the lower and higher ends of the spectrum.}
\label{unlensed_spectra}
\end{figure*}
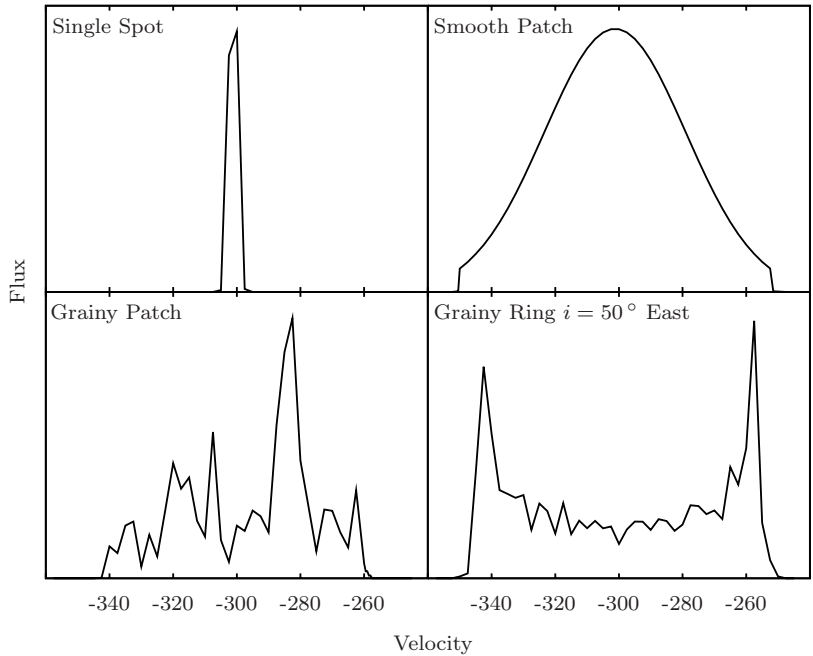

\begin{figure*} 
\centering    
\subfigure[Single Spot]{\includegraphics[width=65mm]{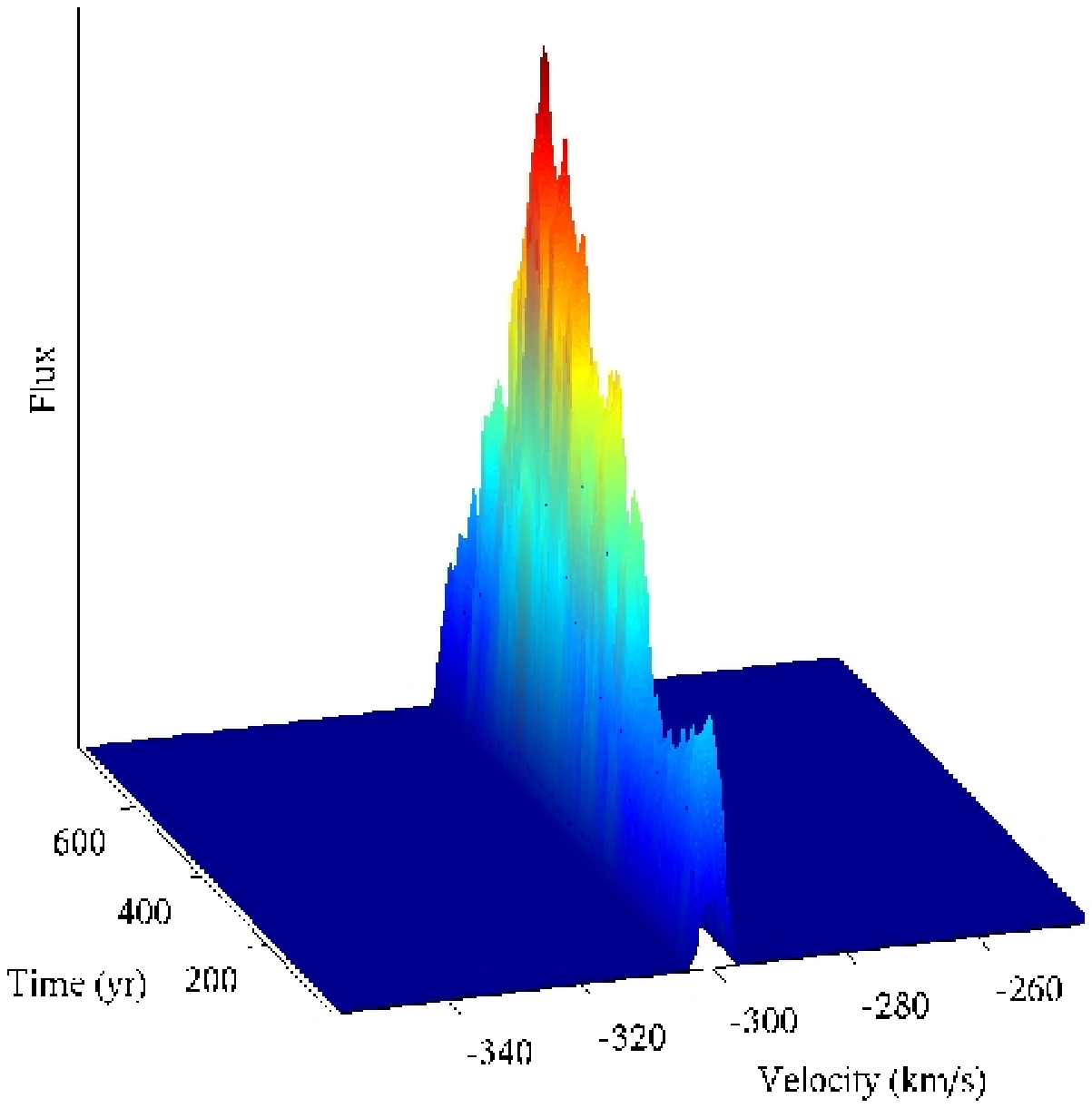}}
\subfigure[Smooth Patch]{\includegraphics[width=65mm]{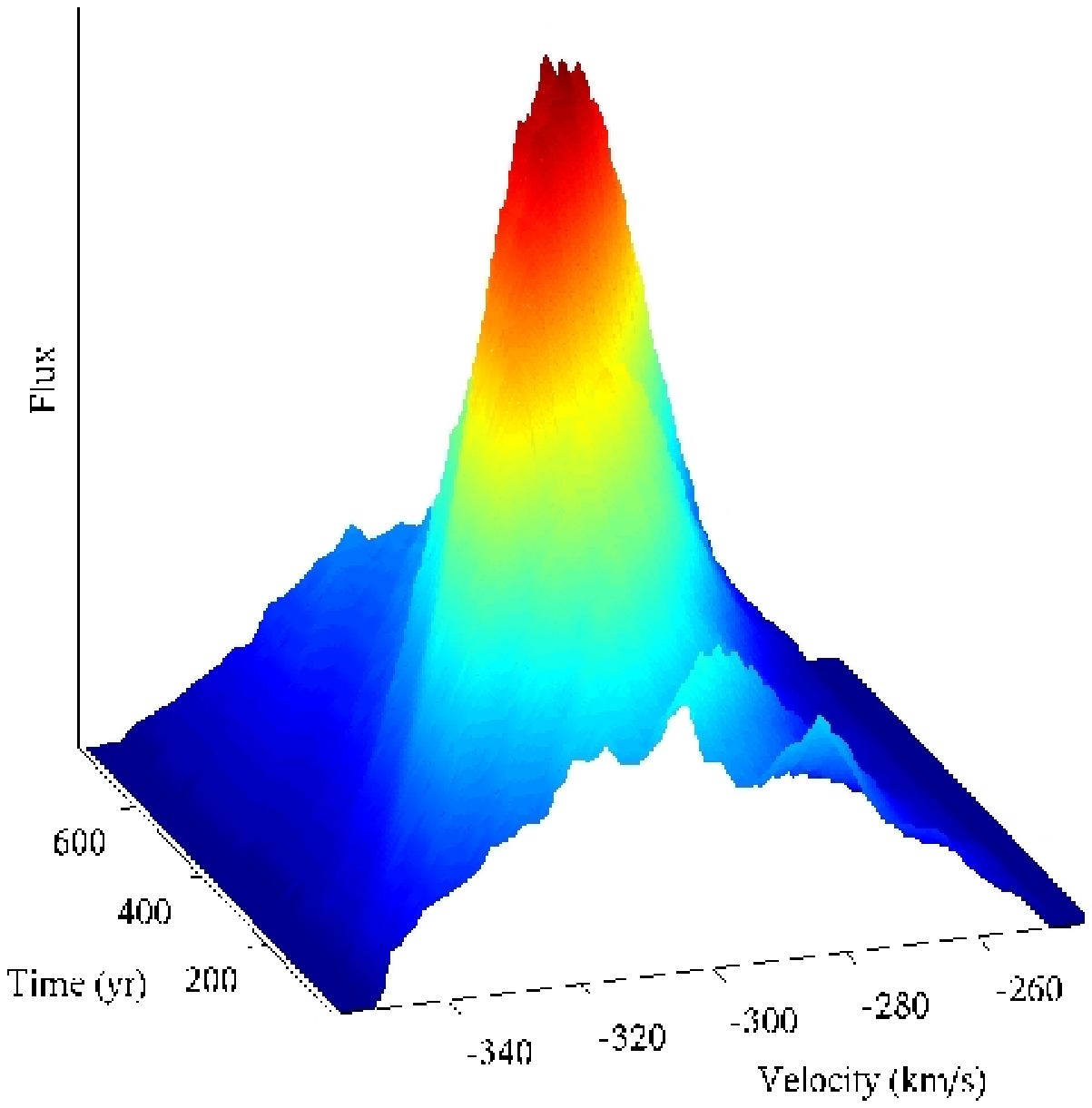}}
\subfigure[Grainy Patch]{\includegraphics[width=65mm]{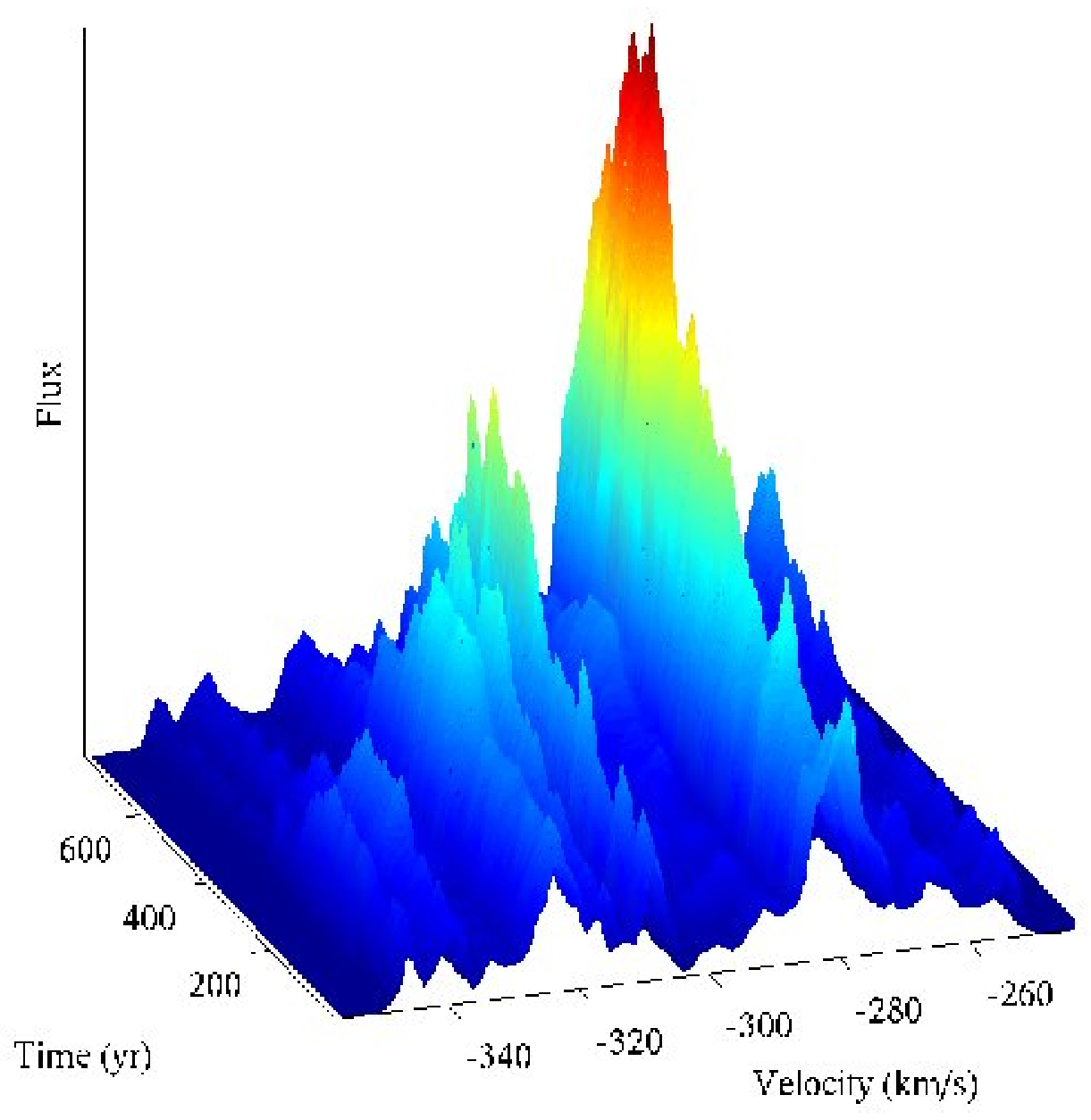}}
\subfigure[Grainy Ring, East, $75\,^{\circ}$]{\includegraphics[width=65mm]{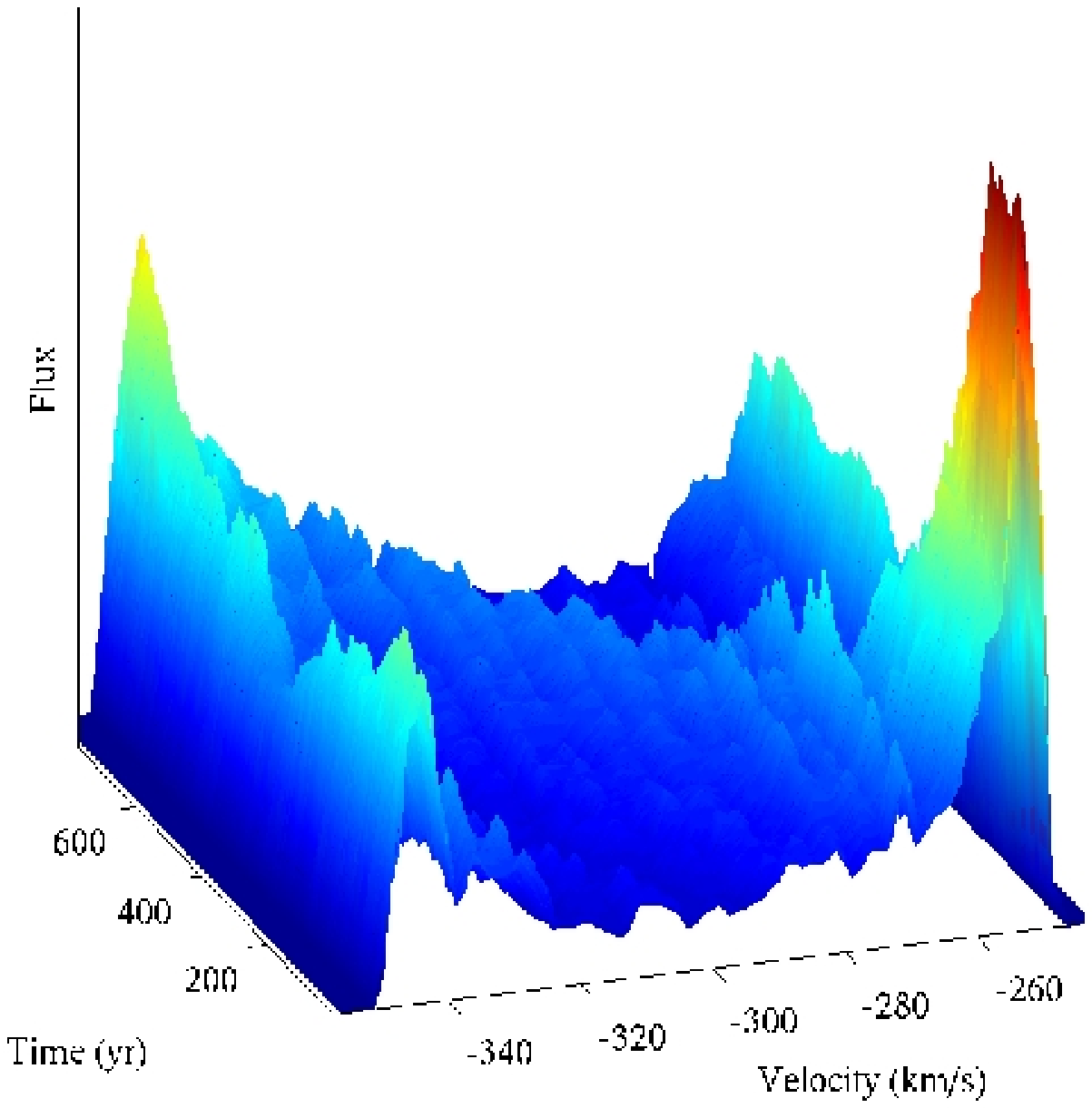}}  
\caption{Spectra of the four simplest source models after lensing within image A1, corresponding in the same order to the spectra in Figure \ref{unlensed_spectra}. As the source moves behind the lens 
a spectrum is captured at each point, these are concatenated to produce a continuous spectrum over time, so that the change can be seen. The Velocity axis is the velocity present in the source model. The Flux is the model flux. The Time axis indicates the passing of time as the source moves from an arbitrary zero point along
a path we have chosen; \shl a Time of 600 years corresponds to a source path distance of 0.1 pc (7.8 ER). \ehl For each model the spectrum
doesn't not change radically; the overall shape is retained but small peaks may come and go. The Grainy Patch shows
the most variability on statistical measures (Figure \ref{overall}).}
\label{lensed_spectra}
\end{figure*}

\begin{figure*}
\centering    
\begingroup%
\makeatletter%
\newcommand{\GNUPLOTspecial}{%
  \@sanitize\catcode`\%=14\relax\special}%
\setlength{\unitlength}{0.0500bp}%
\begin{picture}(5760,4320)(0,0)%
  {\GNUPLOTspecial{"
/gnudict 256 dict def
gnudict begin
%
%
/Color false def
/Blacktext true def
/Solid false def
/Dashlength 1 def
/Landscape false def
/Level1 false def
/Rounded false def
/ClipToBoundingBox false def
/TransparentPatterns false def
/gnulinewidth 7.000 def
/userlinewidth gnulinewidth def
/Gamma 1.0 def
/vshift -66 def
/dl1 {
  10.0 Dashlength mul mul
  Rounded { currentlinewidth 0.75 mul sub dup 0 le { pop 0.01 } if } if
} def
/dl2 {
  10.0 Dashlength mul mul
  Rounded { currentlinewidth 0.75 mul add } if
} def
/hpt_ 31.5 def
/vpt_ 31.5 def
/hpt hpt_ def
/vpt vpt_ def
Level1 {} {
/SDict 10 dict def
systemdict /pdfmark known not {
  userdict /pdfmark systemdict /cleartomark get put
} if
SDict begin [
  /Title (histogram.tex)
  /Subject (gnuplot plot)
  /Creator (gnuplot 4.4 patchlevel 0)
  /Author (hgarsden)
  /CreationDate (Wed Oct  6 19:02:02 2010)
  /DOCINFO pdfmark
end
} ifelse
/doclip {
  ClipToBoundingBox {
    newpath 0 0 moveto 288 0 lineto 288 216 lineto 0 216 lineto closepath
    clip
  } if
} def
%
%
%
/M {moveto} bind def
/L {lineto} bind def
/R {rmoveto} bind def
/V {rlineto} bind def
/N {newpath moveto} bind def
/Z {closepath} bind def
/C {setrgbcolor} bind def
/f {rlineto fill} bind def
/Gshow {show} def   
/vpt2 vpt 2 mul def
/hpt2 hpt 2 mul def
/Lshow {currentpoint stroke M 0 vshift R 
	Blacktext {gsave 0 setgray show grestore} {show} ifelse} def
/Rshow {currentpoint stroke M dup stringwidth pop neg vshift R
	Blacktext {gsave 0 setgray show grestore} {show} ifelse} def
/Cshow {currentpoint stroke M dup stringwidth pop -2 div vshift R 
	Blacktext {gsave 0 setgray show grestore} {show} ifelse} def
/UP {dup vpt_ mul /vpt exch def hpt_ mul /hpt exch def
  /hpt2 hpt 2 mul def /vpt2 vpt 2 mul def} def
/DL {Color {setrgbcolor Solid {pop []} if 0 setdash}
 {pop pop pop 0 setgray Solid {pop []} if 0 setdash} ifelse} def
/BL {stroke userlinewidth 2 mul setlinewidth
	Rounded {1 setlinejoin 1 setlinecap} if} def
/AL {stroke userlinewidth 2 div setlinewidth
	Rounded {1 setlinejoin 1 setlinecap} if} def
/UL {dup gnulinewidth mul /userlinewidth exch def
	dup 1 lt {pop 1} if 10 mul /udl exch def} def
/PL {stroke userlinewidth setlinewidth
	Rounded {1 setlinejoin 1 setlinecap} if} def
/LCw {1 1 1} def
/LCb {0 0 0} def
/LCa {0 0 0} def
/LC0 {1 0 0} def
/LC1 {0 1 0} def
/LC2 {0 0 1} def
/LC3 {1 0 1} def
/LC4 {0 1 1} def
/LC5 {1 1 0} def
/LC6 {0 0 0} def
/LC7 {1 0.3 0} def
/LC8 {0.5 0.5 0.5} def
/LTw {PL [] 1 setgray} def
/LTb {BL [] LCb DL} def
/LTa {AL [1 udl mul 2 udl mul] 0 setdash LCa setrgbcolor} def
/LT0 {PL [] LC0 DL} def
/LT1 {PL [4 dl1 2 dl2] LC1 DL} def
/LT2 {PL [2 dl1 3 dl2] LC2 DL} def
/LT3 {PL [1 dl1 1.5 dl2] LC3 DL} def
/LT4 {PL [6 dl1 2 dl2 1 dl1 2 dl2] LC4 DL} def
/LT5 {PL [3 dl1 3 dl2 1 dl1 3 dl2] LC5 DL} def
/LT6 {PL [2 dl1 2 dl2 2 dl1 6 dl2] LC6 DL} def
/LT7 {PL [1 dl1 2 dl2 6 dl1 2 dl2 1 dl1 2 dl2] LC7 DL} def
/LT8 {PL [2 dl1 2 dl2 2 dl1 2 dl2 2 dl1 2 dl2 2 dl1 4 dl2] LC8 DL} def
/Pnt {stroke [] 0 setdash gsave 1 setlinecap M 0 0 V stroke grestore} def
/Dia {stroke [] 0 setdash 2 copy vpt add M
  hpt neg vpt neg V hpt vpt neg V
  hpt vpt V hpt neg vpt V closepath stroke
  Pnt} def
/Pls {stroke [] 0 setdash vpt sub M 0 vpt2 V
  currentpoint stroke M
  hpt neg vpt neg R hpt2 0 V stroke
 } def
/Box {stroke [] 0 setdash 2 copy exch hpt sub exch vpt add M
  0 vpt2 neg V hpt2 0 V 0 vpt2 V
  hpt2 neg 0 V closepath stroke
  Pnt} def
/Crs {stroke [] 0 setdash exch hpt sub exch vpt add M
  hpt2 vpt2 neg V currentpoint stroke M
  hpt2 neg 0 R hpt2 vpt2 V stroke} def
/TriU {stroke [] 0 setdash 2 copy vpt 1.12 mul add M
  hpt neg vpt -1.62 mul V
  hpt 2 mul 0 V
  hpt neg vpt 1.62 mul V closepath stroke
  Pnt} def
/Star {2 copy Pls Crs} def
/BoxF {stroke [] 0 setdash exch hpt sub exch vpt add M
  0 vpt2 neg V hpt2 0 V 0 vpt2 V
  hpt2 neg 0 V closepath fill} def
/TriUF {stroke [] 0 setdash vpt 1.12 mul add M
  hpt neg vpt -1.62 mul V
  hpt 2 mul 0 V
  hpt neg vpt 1.62 mul V closepath fill} def
/TriD {stroke [] 0 setdash 2 copy vpt 1.12 mul sub M
  hpt neg vpt 1.62 mul V
  hpt 2 mul 0 V
  hpt neg vpt -1.62 mul V closepath stroke
  Pnt} def
/TriDF {stroke [] 0 setdash vpt 1.12 mul sub M
  hpt neg vpt 1.62 mul V
  hpt 2 mul 0 V
  hpt neg vpt -1.62 mul V closepath fill} def
/DiaF {stroke [] 0 setdash vpt add M
  hpt neg vpt neg V hpt vpt neg V
  hpt vpt V hpt neg vpt V closepath fill} def
/Pent {stroke [] 0 setdash 2 copy gsave
  translate 0 hpt M 4 {72 rotate 0 hpt L} repeat
  closepath stroke grestore Pnt} def
/PentF {stroke [] 0 setdash gsave
  translate 0 hpt M 4 {72 rotate 0 hpt L} repeat
  closepath fill grestore} def
/Circle {stroke [] 0 setdash 2 copy
  hpt 0 360 arc stroke Pnt} def
/CircleF {stroke [] 0 setdash hpt 0 360 arc fill} def
/C0 {BL [] 0 setdash 2 copy moveto vpt 90 450 arc} bind def
/C1 {BL [] 0 setdash 2 copy moveto
	2 copy vpt 0 90 arc closepath fill
	vpt 0 360 arc closepath} bind def
/C2 {BL [] 0 setdash 2 copy moveto
	2 copy vpt 90 180 arc closepath fill
	vpt 0 360 arc closepath} bind def
/C3 {BL [] 0 setdash 2 copy moveto
	2 copy vpt 0 180 arc closepath fill
	vpt 0 360 arc closepath} bind def
/C4 {BL [] 0 setdash 2 copy moveto
	2 copy vpt 180 270 arc closepath fill
	vpt 0 360 arc closepath} bind def
/C5 {BL [] 0 setdash 2 copy moveto
	2 copy vpt 0 90 arc
	2 copy moveto
	2 copy vpt 180 270 arc closepath fill
	vpt 0 360 arc} bind def
/C6 {BL [] 0 setdash 2 copy moveto
	2 copy vpt 90 270 arc closepath fill
	vpt 0 360 arc closepath} bind def
/C7 {BL [] 0 setdash 2 copy moveto
	2 copy vpt 0 270 arc closepath fill
	vpt 0 360 arc closepath} bind def
/C8 {BL [] 0 setdash 2 copy moveto
	2 copy vpt 270 360 arc closepath fill
	vpt 0 360 arc closepath} bind def
/C9 {BL [] 0 setdash 2 copy moveto
	2 copy vpt 270 450 arc closepath fill
	vpt 0 360 arc closepath} bind def
/C10 {BL [] 0 setdash 2 copy 2 copy moveto vpt 270 360 arc closepath fill
	2 copy moveto
	2 copy vpt 90 180 arc closepath fill
	vpt 0 360 arc closepath} bind def
/C11 {BL [] 0 setdash 2 copy moveto
	2 copy vpt 0 180 arc closepath fill
	2 copy moveto
	2 copy vpt 270 360 arc closepath fill
	vpt 0 360 arc closepath} bind def
/C12 {BL [] 0 setdash 2 copy moveto
	2 copy vpt 180 360 arc closepath fill
	vpt 0 360 arc closepath} bind def
/C13 {BL [] 0 setdash 2 copy moveto
	2 copy vpt 0 90 arc closepath fill
	2 copy moveto
	2 copy vpt 180 360 arc closepath fill
	vpt 0 360 arc closepath} bind def
/C14 {BL [] 0 setdash 2 copy moveto
	2 copy vpt 90 360 arc closepath fill
	vpt 0 360 arc} bind def
/C15 {BL [] 0 setdash 2 copy vpt 0 360 arc closepath fill
	vpt 0 360 arc closepath} bind def
/Rec {newpath 4 2 roll moveto 1 index 0 rlineto 0 exch rlineto
	neg 0 rlineto closepath} bind def
/Square {dup Rec} bind def
/Bsquare {vpt sub exch vpt sub exch vpt2 Square} bind def
/S0 {BL [] 0 setdash 2 copy moveto 0 vpt rlineto BL Bsquare} bind def
/S1 {BL [] 0 setdash 2 copy vpt Square fill Bsquare} bind def
/S2 {BL [] 0 setdash 2 copy exch vpt sub exch vpt Square fill Bsquare} bind def
/S3 {BL [] 0 setdash 2 copy exch vpt sub exch vpt2 vpt Rec fill Bsquare} bind def
/S4 {BL [] 0 setdash 2 copy exch vpt sub exch vpt sub vpt Square fill Bsquare} bind def
/S5 {BL [] 0 setdash 2 copy 2 copy vpt Square fill
	exch vpt sub exch vpt sub vpt Square fill Bsquare} bind def
/S6 {BL [] 0 setdash 2 copy exch vpt sub exch vpt sub vpt vpt2 Rec fill Bsquare} bind def
/S7 {BL [] 0 setdash 2 copy exch vpt sub exch vpt sub vpt vpt2 Rec fill
	2 copy vpt Square fill Bsquare} bind def
/S8 {BL [] 0 setdash 2 copy vpt sub vpt Square fill Bsquare} bind def
/S9 {BL [] 0 setdash 2 copy vpt sub vpt vpt2 Rec fill Bsquare} bind def
/S10 {BL [] 0 setdash 2 copy vpt sub vpt Square fill 2 copy exch vpt sub exch vpt Square fill
	Bsquare} bind def
/S11 {BL [] 0 setdash 2 copy vpt sub vpt Square fill 2 copy exch vpt sub exch vpt2 vpt Rec fill
	Bsquare} bind def
/S12 {BL [] 0 setdash 2 copy exch vpt sub exch vpt sub vpt2 vpt Rec fill Bsquare} bind def
/S13 {BL [] 0 setdash 2 copy exch vpt sub exch vpt sub vpt2 vpt Rec fill
	2 copy vpt Square fill Bsquare} bind def
/S14 {BL [] 0 setdash 2 copy exch vpt sub exch vpt sub vpt2 vpt Rec fill
	2 copy exch vpt sub exch vpt Square fill Bsquare} bind def
/S15 {BL [] 0 setdash 2 copy Bsquare fill Bsquare} bind def
/D0 {gsave translate 45 rotate 0 0 S0 stroke grestore} bind def
/D1 {gsave translate 45 rotate 0 0 S1 stroke grestore} bind def
/D2 {gsave translate 45 rotate 0 0 S2 stroke grestore} bind def
/D3 {gsave translate 45 rotate 0 0 S3 stroke grestore} bind def
/D4 {gsave translate 45 rotate 0 0 S4 stroke grestore} bind def
/D5 {gsave translate 45 rotate 0 0 S5 stroke grestore} bind def
/D6 {gsave translate 45 rotate 0 0 S6 stroke grestore} bind def
/D7 {gsave translate 45 rotate 0 0 S7 stroke grestore} bind def
/D8 {gsave translate 45 rotate 0 0 S8 stroke grestore} bind def
/D9 {gsave translate 45 rotate 0 0 S9 stroke grestore} bind def
/D10 {gsave translate 45 rotate 0 0 S10 stroke grestore} bind def
/D11 {gsave translate 45 rotate 0 0 S11 stroke grestore} bind def
/D12 {gsave translate 45 rotate 0 0 S12 stroke grestore} bind def
/D13 {gsave translate 45 rotate 0 0 S13 stroke grestore} bind def
/D14 {gsave translate 45 rotate 0 0 S14 stroke grestore} bind def
/D15 {gsave translate 45 rotate 0 0 S15 stroke grestore} bind def
/DiaE {stroke [] 0 setdash vpt add M
  hpt neg vpt neg V hpt vpt neg V
  hpt vpt V hpt neg vpt V closepath stroke} def
/BoxE {stroke [] 0 setdash exch hpt sub exch vpt add M
  0 vpt2 neg V hpt2 0 V 0 vpt2 V
  hpt2 neg 0 V closepath stroke} def
/TriUE {stroke [] 0 setdash vpt 1.12 mul add M
  hpt neg vpt -1.62 mul V
  hpt 2 mul 0 V
  hpt neg vpt 1.62 mul V closepath stroke} def
/TriDE {stroke [] 0 setdash vpt 1.12 mul sub M
  hpt neg vpt 1.62 mul V
  hpt 2 mul 0 V
  hpt neg vpt -1.62 mul V closepath stroke} def
/PentE {stroke [] 0 setdash gsave
  translate 0 hpt M 4 {72 rotate 0 hpt L} repeat
  closepath stroke grestore} def
/CircE {stroke [] 0 setdash 
  hpt 0 360 arc stroke} def
/Opaque {gsave closepath 1 setgray fill grestore 0 setgray closepath} def
/DiaW {stroke [] 0 setdash vpt add M
  hpt neg vpt neg V hpt vpt neg V
  hpt vpt V hpt neg vpt V Opaque stroke} def
/BoxW {stroke [] 0 setdash exch hpt sub exch vpt add M
  0 vpt2 neg V hpt2 0 V 0 vpt2 V
  hpt2 neg 0 V Opaque stroke} def
/TriUW {stroke [] 0 setdash vpt 1.12 mul add M
  hpt neg vpt -1.62 mul V
  hpt 2 mul 0 V
  hpt neg vpt 1.62 mul V Opaque stroke} def
/TriDW {stroke [] 0 setdash vpt 1.12 mul sub M
  hpt neg vpt 1.62 mul V
  hpt 2 mul 0 V
  hpt neg vpt -1.62 mul V Opaque stroke} def
/PentW {stroke [] 0 setdash gsave
  translate 0 hpt M 4 {72 rotate 0 hpt L} repeat
  Opaque stroke grestore} def
/CircW {stroke [] 0 setdash 
  hpt 0 360 arc Opaque stroke} def
/BoxFill {gsave Rec 1 setgray fill grestore} def
/Density {
  /Fillden exch def
  currentrgbcolor
  /ColB exch def /ColG exch def /ColR exch def
  /ColR ColR Fillden mul Fillden sub 1 add def
  /ColG ColG Fillden mul Fillden sub 1 add def
  /ColB ColB Fillden mul Fillden sub 1 add def
  ColR ColG ColB setrgbcolor} def
/BoxColFill {gsave Rec PolyFill} def
/PolyFill {gsave Density fill grestore grestore} def
/h {rlineto rlineto rlineto gsave closepath fill grestore} bind def
%
%
/PatternFill {gsave /PFa [ 9 2 roll ] def
  PFa 0 get PFa 2 get 2 div add PFa 1 get PFa 3 get 2 div add translate
  PFa 2 get -2 div PFa 3 get -2 div PFa 2 get PFa 3 get Rec
  gsave 1 setgray fill grestore clip
  currentlinewidth 0.5 mul setlinewidth
  /PFs PFa 2 get dup mul PFa 3 get dup mul add sqrt def
  0 0 M PFa 5 get rotate PFs -2 div dup translate
  0 1 PFs PFa 4 get div 1 add floor cvi
	{PFa 4 get mul 0 M 0 PFs V} for
  0 PFa 6 get ne {
	0 1 PFs PFa 4 get div 1 add floor cvi
	{PFa 4 get mul 0 2 1 roll M PFs 0 V} for
 } if
  stroke grestore} def
/languagelevel where
 {pop languagelevel} {1} ifelse
 2 lt
	{/InterpretLevel1 true def}
	{/InterpretLevel1 Level1 def}
 ifelse
%
%
/Level2PatternFill {
/Tile8x8 {/PaintType 2 /PatternType 1 /TilingType 1 /BBox [0 0 8 8] /XStep 8 /YStep 8}
	bind def
/KeepColor {currentrgbcolor [/Pattern /DeviceRGB] setcolorspace} bind def
<< Tile8x8
 /PaintProc {0.5 setlinewidth pop 0 0 M 8 8 L 0 8 M 8 0 L stroke} 
>> matrix makepattern
/Pat1 exch def
<< Tile8x8
 /PaintProc {0.5 setlinewidth pop 0 0 M 8 8 L 0 8 M 8 0 L stroke
	0 4 M 4 8 L 8 4 L 4 0 L 0 4 L stroke}
>> matrix makepattern
/Pat2 exch def
<< Tile8x8
 /PaintProc {0.5 setlinewidth pop 0 0 M 0 8 L
	8 8 L 8 0 L 0 0 L fill}
>> matrix makepattern
/Pat3 exch def
<< Tile8x8
 /PaintProc {0.5 setlinewidth pop -4 8 M 8 -4 L
	0 12 M 12 0 L stroke}
>> matrix makepattern
/Pat4 exch def
<< Tile8x8
 /PaintProc {0.5 setlinewidth pop -4 0 M 8 12 L
	0 -4 M 12 8 L stroke}
>> matrix makepattern
/Pat5 exch def
<< Tile8x8
 /PaintProc {0.5 setlinewidth pop -2 8 M 4 -4 L
	0 12 M 8 -4 L 4 12 M 10 0 L stroke}
>> matrix makepattern
/Pat6 exch def
<< Tile8x8
 /PaintProc {0.5 setlinewidth pop -2 0 M 4 12 L
	0 -4 M 8 12 L 4 -4 M 10 8 L stroke}
>> matrix makepattern
/Pat7 exch def
<< Tile8x8
 /PaintProc {0.5 setlinewidth pop 8 -2 M -4 4 L
	12 0 M -4 8 L 12 4 M 0 10 L stroke}
>> matrix makepattern
/Pat8 exch def
<< Tile8x8
 /PaintProc {0.5 setlinewidth pop 0 -2 M 12 4 L
	-4 0 M 12 8 L -4 4 M 8 10 L stroke}
>> matrix makepattern
/Pat9 exch def
/Pattern1 {PatternBgnd KeepColor Pat1 setpattern} bind def
/Pattern2 {PatternBgnd KeepColor Pat2 setpattern} bind def
/Pattern3 {PatternBgnd KeepColor Pat3 setpattern} bind def
/Pattern4 {PatternBgnd KeepColor Landscape {Pat5} {Pat4} ifelse setpattern} bind def
/Pattern5 {PatternBgnd KeepColor Landscape {Pat4} {Pat5} ifelse setpattern} bind def
/Pattern6 {PatternBgnd KeepColor Landscape {Pat9} {Pat6} ifelse setpattern} bind def
/Pattern7 {PatternBgnd KeepColor Landscape {Pat8} {Pat7} ifelse setpattern} bind def
} def
%
%
%
/PatternBgnd {
  TransparentPatterns {} {gsave 1 setgray fill grestore} ifelse
} def
%
%
/Level1PatternFill {
/Pattern1 {0.250 Density} bind def
/Pattern2 {0.500 Density} bind def
/Pattern3 {0.750 Density} bind def
/Pattern4 {0.125 Density} bind def
/Pattern5 {0.375 Density} bind def
/Pattern6 {0.625 Density} bind def
/Pattern7 {0.875 Density} bind def
} def
%
%
Level1 {Level1PatternFill} {Level2PatternFill} ifelse
/Symbol-Oblique /Symbol findfont [1 0 .167 1 0 0] makefont
dup length dict begin {1 index /FID eq {pop pop} {def} ifelse} forall
currentdict end definefont pop
end
gnudict begin
gsave
doclip
0 0 translate
0.050 0.050 scale
0 setgray
newpath
1.000 UL
LTb
0 2160 M
0 37 V
0 4319 M
0 -37 V
480 2160 M
0 37 V
0 2122 R
0 -37 V
960 2160 M
0 37 V
0 2122 R
0 -37 V
1440 2160 M
0 37 V
0 2122 R
0 -37 V
1919 2160 M
0 37 V
0 2122 R
0 -37 V
2399 2160 M
0 37 V
0 2122 R
0 -37 V
2879 2160 M
0 37 V
0 2122 R
0 -37 V
stroke
0 4319 N
0 2160 L
2879 0 V
0 2159 V
0 4319 L
Z stroke
LCb setrgbcolor
LTb
1.000 UP
1.000 UL
LTb
LCb setrgbcolor
LTb
1.000 UL
LTb
0 2160 M
15 0 V
75 0 V
75 0 V
45 59 V
15 529 V
15 1057 V
75 274 V
75 -274 V
75 274 V
75 -274 V
75 0 V
75 274 V
75 -274 V
75 274 V
75 -274 V
75 274 V
75 -274 V
75 274 V
75 -274 V
75 274 V
75 -274 V
75 274 V
74 0 V
75 -274 V
75 274 V
75 -274 V
75 274 V
75 -274 V
75 274 V
75 -274 V
75 274 V
75 -274 V
75 0 V
75 274 V
75 -274 V
75 274 V
75 -274 V
75 274 V
15 -1331 V
15 -529 V
45 -59 V
75 0 V
75 0 V
15 0 V
stroke
0 4319 N
0 2160 L
2879 0 V
0 2159 V
0 4319 L
Z stroke
1.000 UP
1.000 UL
LTb
1.000 UL
LTb
2880 2160 M
0 37 V
0 2122 R
0 -37 V
3360 2160 M
0 37 V
0 2122 R
0 -37 V
3840 2160 M
0 37 V
0 2122 R
0 -37 V
4320 2160 M
0 37 V
0 2122 R
0 -37 V
4799 2160 M
0 37 V
0 2122 R
0 -37 V
5279 2160 M
0 37 V
0 2122 R
0 -37 V
5759 2160 M
0 37 V
0 2122 R
0 -37 V
-2879 37 R
0 -2159 R
2879 0 R
0 2159 V
-2879 0 V
1.000 UP
stroke
LCb setrgbcolor
LTb
1.000 UL
LTb
2880 2160 M
15 0 V
75 0 V
75 0 V
75 0 V
75 0 V
75 17 V
75 101 V
75 201 V
75 255 V
75 284 V
75 293 V
75 283 V
75 247 V
75 161 V
75 46 V
75 8 V
75 26 V
75 32 V
75 18 V
75 0 V
74 -18 V
75 -31 V
75 -27 V
75 -7 V
75 -40 V
75 -153 V
75 -243 V
75 -280 V
75 -293 V
75 -285 V
75 -258 V
75 -206 V
75 -110 V
75 -21 V
75 0 V
75 0 V
75 0 V
75 0 V
75 0 V
15 0 V
2880 4319 M
0 -2159 R
2879 0 R
0 2159 V
-2879 0 V
1.000 UP
0 0 M
0 37 L
0 2160 M
0 -37 V
480 0 M
0 37 V
0 2123 R
0 -37 V
960 0 M
0 37 V
0 2123 R
0 -37 V
1440 0 M
0 37 V
0 2123 R
0 -37 V
1919 0 M
0 37 V
0 2123 R
0 -37 V
2399 0 M
0 37 V
0 2123 R
0 -37 V
2879 0 M
0 37 V
0 2123 R
0 -37 V
0 2160 M
0 0 L
2879 0 L
0 2160 V
0 2160 M
1.000 UP
stroke
LCb setrgbcolor
LTb
1.000 UL
LTb
0 0 M
15 0 L
90 0 L
75 0 V
45 59 V
15 530 V
15 1057 V
75 274 V
75 -274 V
75 274 V
75 -274 V
75 0 V
75 274 V
75 -274 V
75 274 V
75 -274 V
75 274 V
75 -274 V
75 274 V
75 -274 V
75 274 V
75 -274 V
75 274 V
74 0 V
75 -274 V
75 274 V
75 -274 V
75 274 V
75 -274 V
75 274 V
75 -274 V
75 274 V
75 -274 V
75 0 V
75 274 V
75 -274 V
75 274 V
75 -274 V
75 274 V
2654 589 L
2669 59 L
2714 0 L
75 0 V
75 0 V
15 0 V
0 2160 M
0 0 L
2879 0 L
0 2160 V
0 2160 M
1.000 UP
2880 0 M
0 37 V
0 2123 R
0 -37 V
3360 0 M
0 37 V
0 2123 R
0 -37 V
3840 0 M
0 37 V
0 2123 R
0 -37 V
4320 0 M
0 37 V
0 2123 R
0 -37 V
4799 0 M
0 37 V
0 2123 R
0 -37 V
5279 0 M
0 37 V
0 2123 R
0 -37 V
5759 0 M
0 37 V
0 2123 R
0 -37 V
-2879 37 R
2880 0 M
5759 0 L
0 2160 V
-2879 0 V
stroke
LCb setrgbcolor
LTb
1.000 UP
1.000 UL
LTb
LCb setrgbcolor
LTb
1.000 UL
LTb
2880 0 M
15 0 V
75 0 V
75 0 V
75 0 V
75 0 V
75 22 V
75 134 V
75 252 V
75 247 V
75 208 V
75 206 V
75 247 V
75 297 V
75 217 V
75 62 V
75 9 V
75 34 V
75 42 V
75 19 V
75 0 V
74 -19 V
75 -42 V
75 -33 V
75 -9 V
75 -54 V
75 -206 V
75 -296 V
75 -252 V
75 -207 V
75 -206 V
75 -243 V
75 -255 V
5294 28 L
5369 0 L
75 0 V
75 0 V
75 0 V
75 0 V
75 0 V
15 0 V
2880 2160 M
2880 0 M
5759 0 L
0 2160 V
-2879 0 V
1.000 UP
stroke
grestore
end
showpage
  }}%
  \put(3591,1997){\makebox(0,0){\strut{}Hollow Jet $i=50\,^\circ$}}%
  \put(2919,-499){\makebox(0,0){\strut{}Velocity}}%
  \put(5279,-200){\makebox(0,0){\strut{}40}}%
  \put(4799,-200){\makebox(0,0){\strut{}20}}%
  \put(4320,-200){\makebox(0,0){\strut{}0}}%
  \put(3840,-200){\makebox(0,0){\strut{}-20}}%
  \put(3360,-200){\makebox(0,0){\strut{}-40}}%
  \put(691,1997){\makebox(0,0){\strut{}Hollow Jet $i=0\,^\circ$}}%
  \put(2399,-200){\makebox(0,0){\strut{}40}}%
  \put(1919,-200){\makebox(0,0){\strut{}20}}%
  \put(1440,-200){\makebox(0,0){\strut{}0}}%
  \put(960,-200){\makebox(0,0){\strut{}-20}}%
  \put(480,-200){\makebox(0,0){\strut{}-40}}%
  \put(3541,4156){\makebox(0,0){\strut{}Solid Jet $i=50\,^\circ$}}%
  \put(631,4156){\makebox(0,0){\strut{}Solid Jet $i=0\,^\circ$}}%
  \put(-200,2239){%
  \makebox(0,0){\strut{}Flux}%
  }%
\end{picture}%
\endgroup
\vspace{10mm}
\caption{The spectra of the Solid Jets and Hollow Jets models without lensing. Reading left to right, top to bottom,  the spectra
are for the models: Solid Jet $i=0$, Solid Jet $i=50\,^{\circ}$ South-East, Hollow Jet $i=0$, Hollow Jet $i=50\,^{\circ}$ South-East. The
Solid vs. Hollow models produce very similar spectra, because of the way the models are defined. Both of them change when the 
orientation is changed. A face-on ($i=0$) inclination (left panels) produces a spectrum with an overall square-wave shape (to the
resolution of the velocity slices), but when inclined (right panels)
this transforms to a single peak, for both models. Only the inclination of i=$50\,^{\circ}$ South-East
is shown here as a sample, the peaks get sharper with more inclination. One slight difference is that the Hollow Jet produces a sharper peak than the Solid Jets.}
\label{unlensed_spectra_jet}
\end{figure*}
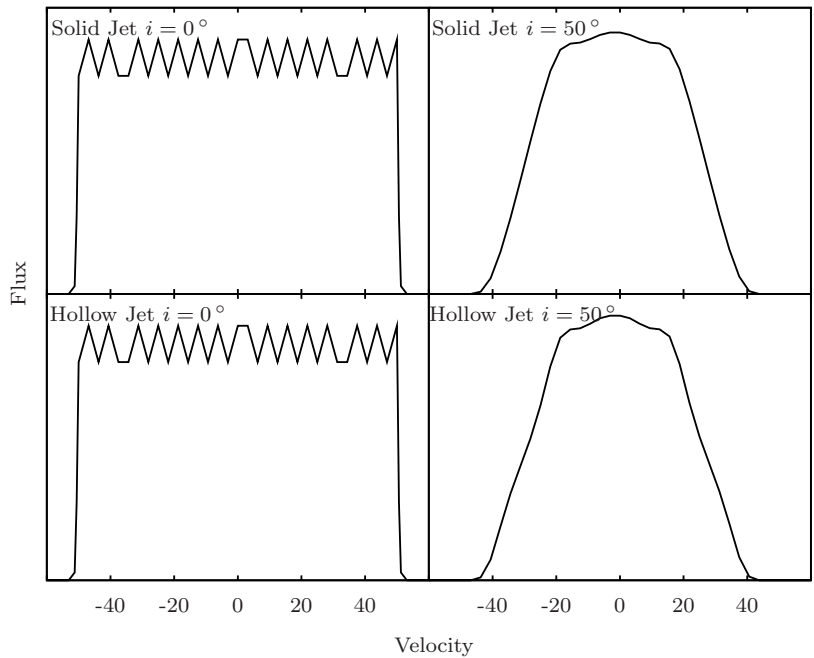

\begin{figure*}
\centering    
\subfigure[Solid Jets, $i=0$]{\includegraphics[width=65mm]{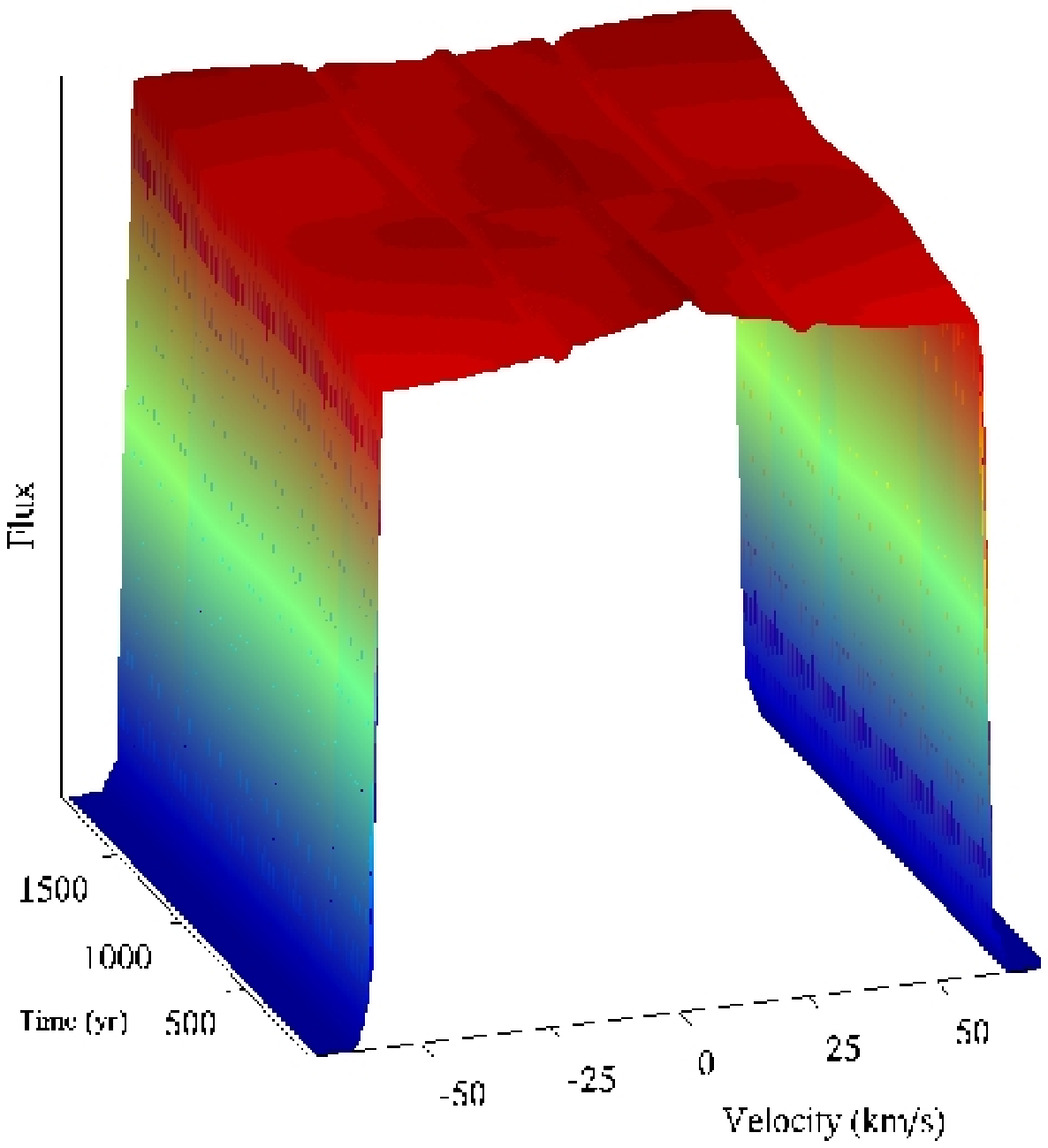}}
\subfigure[Solid Jets, $i=50\,^{\circ}$ South-East]{\includegraphics[width=65mm]{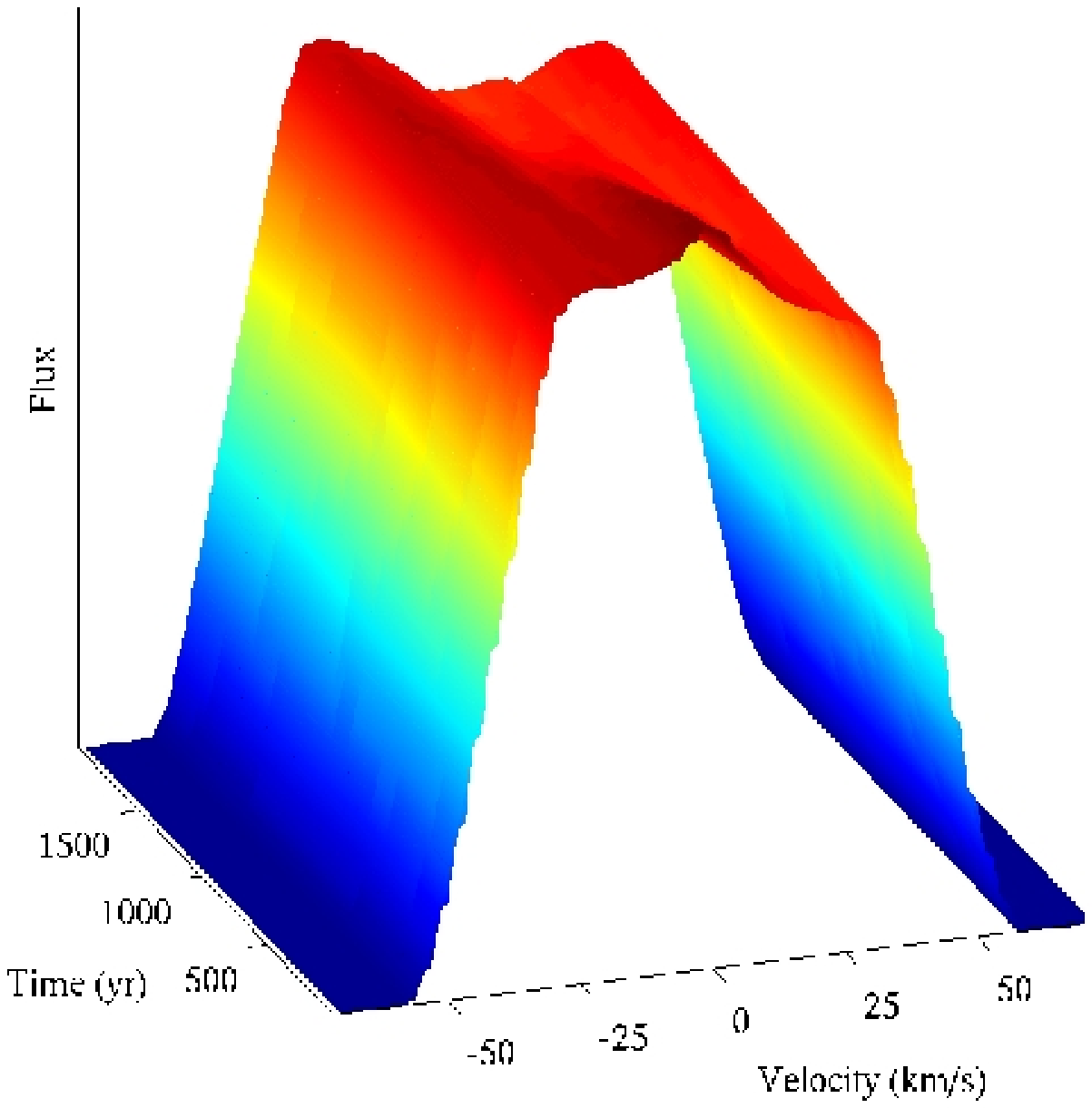}}
\subfigure[Hollow Jets, $i=0$]{\includegraphics[width=65mm]{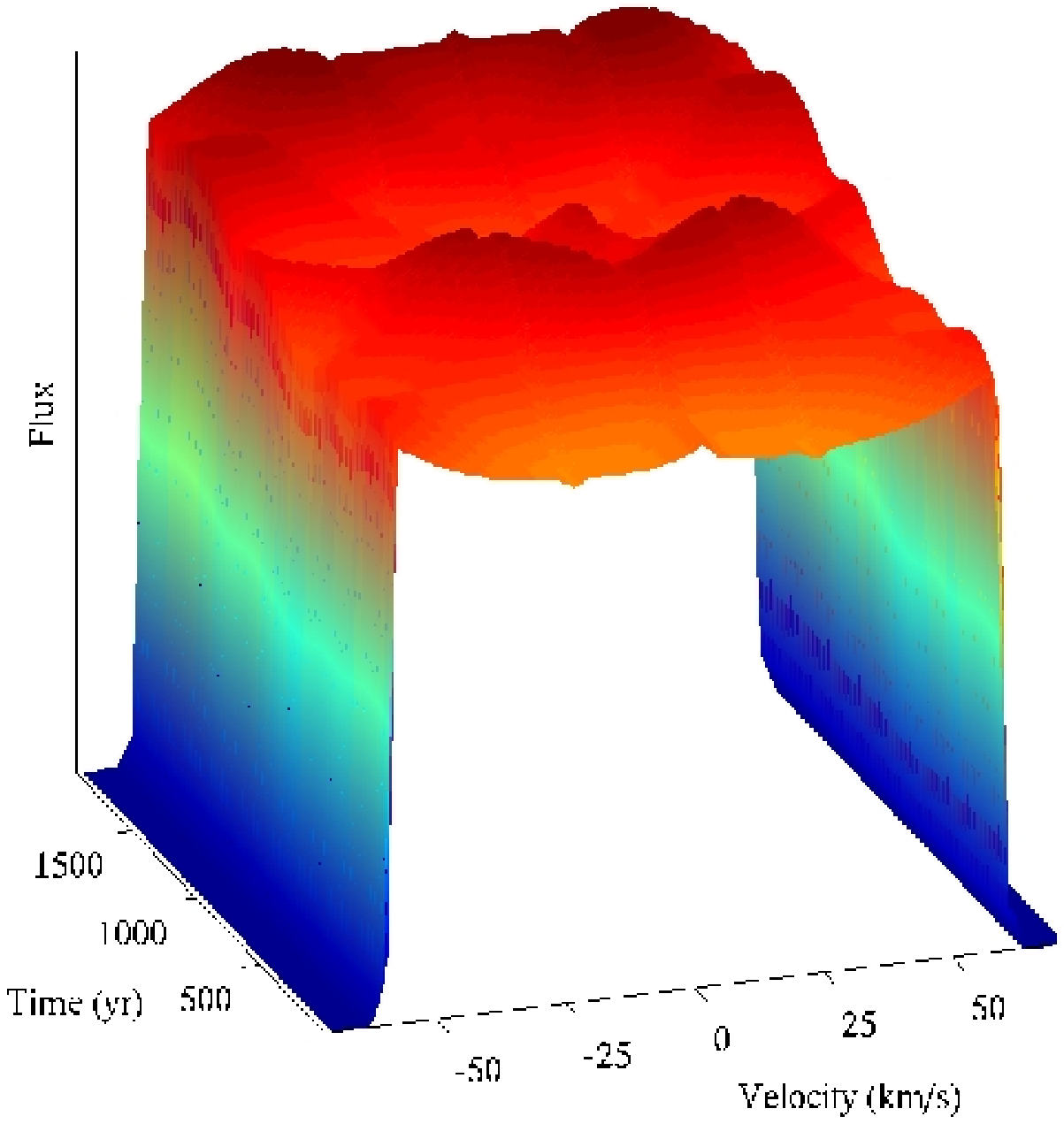}}  
\subfigure[Hollow Jets, $i=50\,^{\circ}$ South-East]{\includegraphics[width=65mm]{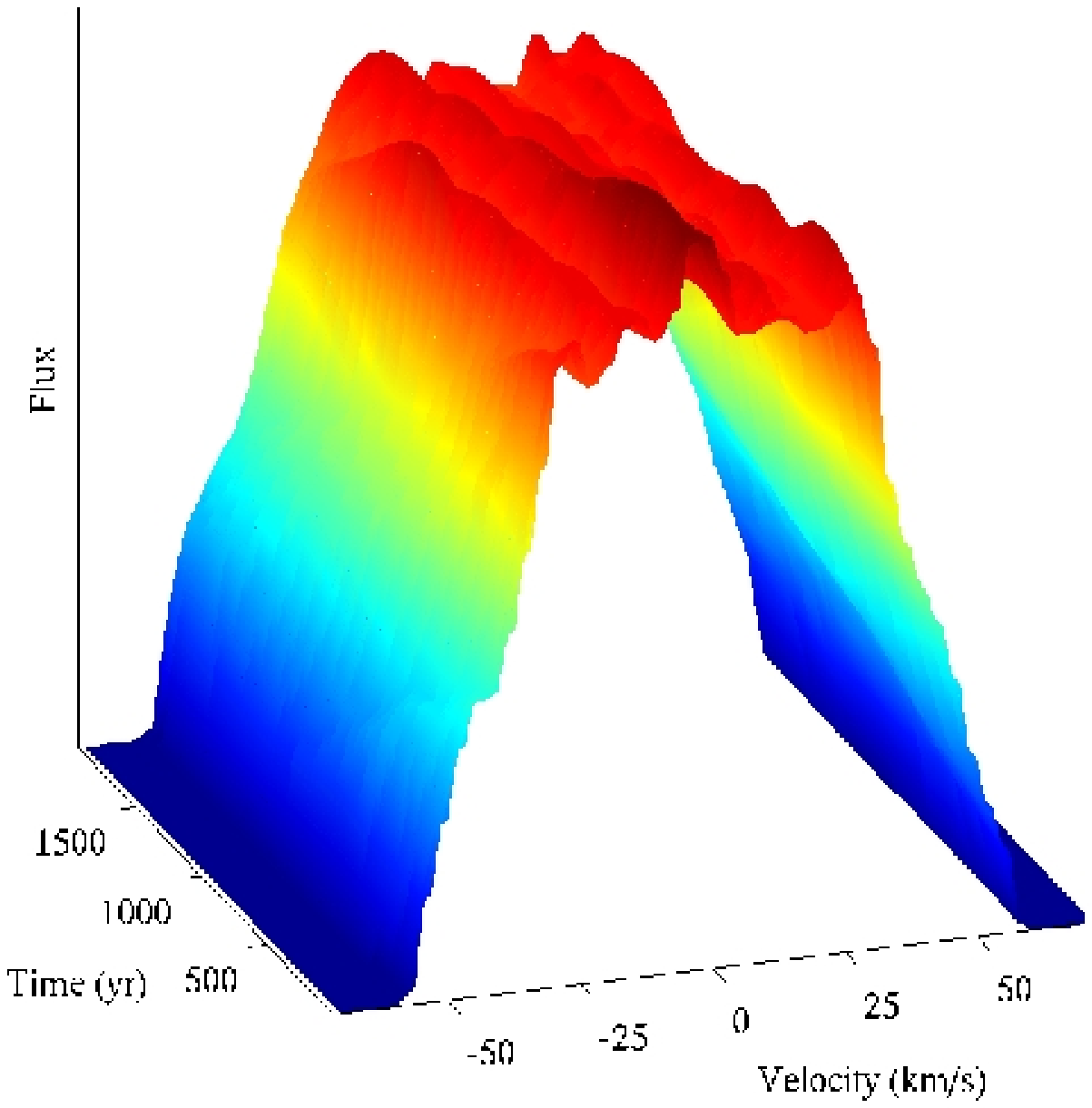}} 
\caption{Spectra of the Jets source models, for  image A1,  after lensing. Each image corresponds
to the spectrum in Figure \ref{unlensed_spectra_jet}. As the source moves behind the lens 
a spectrum is captured at each point, these are concatenated to produce a continuous spectrum over time,  allowing the changes to be seen. The Velocity axis is the velocity present in the source model. The Flux is the model flux. The Time axis indicates the passing of time as the source moves from an arbitrary 0 point along
a path we have chosen; \shl a Time of 1500 years corresponds to a source path distance of 0.26 pc (19 ER). \ehl The models produce different spectra at different orientations but these are little 
affected by microlensing, and change little over time. }
\label{lensed_spectra_jet}
\end{figure*}

\subsubsection{Two patches of masers}

Some masers lie in a group on the side of a jet \cite[e.g. NGC 1052:][]{ngc1052} with a fairly uniform velocity gradient from one side of the group to the other.  Initially we approach this as an unresolved Smooth Patch, modelled as a 2-D Gaussian with a line-of-sight velocity gradient of -350 to -250 km s$^{-1}$ from left to right. The total width of the patch is set at 0.1 pc, which is consistent with the  size of other maser groups \citep[e.g. Mrk348:][]{mrk348}.
As well as a smooth area, we  use a group of spots, and replace the Smooth Patch
with a collection of 49 spots called a Grainy Patch. The extent of the patch and the velocity profile remains the same,  but it consists of Single
Spots, randomly located within a circular region. Figure \ref{schemes} (a) has the schematic for the Grainy Patch,
and (d) shows the resultant model, the Smooth Patch is not shown.

\subsubsection{A Ring of Masers}

If the jet in
 MG 0414+0534 is mostly face-on then it may be possible to observe masers situated in a ring around the jet.
 Figure \ref{schemes} (b) shows the schematic of such a ring,  where there are 200 Single Spots 
in a ring of diameter 0.5 pc.  The location 
of each spot has been randomly perturbed from the true ring. The ring is assumed to lie $\sim$ 1.4 pc along a 
conical jet, with
an opening angle of $10\,^{\circ}$, allowing the ring to be oriented by orienting the jet. The jet has
two degrees of freedom. Firstly, the angle of inclination determines whether the
jet is perpendicular or parallel to the plane of the sky,  in other words parallel to
the plane of the magnification map.  An inclination of $i=0\,^{\circ}$ means the jet is pointing out of the map towards you, and $i=90\,^{\circ}$ is fully side-on.  Secondly, 
when the jet is inclined it may be pointing in a different compass direction on the  map: South/East etc.
The orientation of the ring has an impact on the lensing, as will
be shown later. The ring at an inclination of $0\,^{\circ}$ is shown
in Figure \ref{schemes}  (e).

The jet material is outflowing, but the maser material in the ring is possibly turbulent and  the velocity 
\sbl is not well known. \ehl
However we will assume an overall bulk maser motion in a similar direction to the jet.
Therefore we   set the velocity of the masers 
to be in the same direction as the jet material they lie next to, but scaled  so
the line-of-sight velocity of the entire model falls between -350 and -250 km s$^{-1}$, the same as the for the patches.
This range applies regardless of how the ring is oriented, except when it is face-on, in which case the line-of-sight velocity is the same around the ring and set to 300 km s$^{-1}$.
We will call this model the Grainy Ring.

\subsubsection{Continuous jet masers}

This model consists of two identical jets in a bi-conical configuration outflowing from a quasar core. Each jet is  1.59 pc long, beginning at a
base of width
0.13 pc with an  opening angle of $25\,^{\circ}$ for 0.53 pc, and then narrowing to $8\,^{\circ}$ for 1.06 pc, producing
a jet diameter of 0.64 pc at the limit of the model. The numbers are chosen because of the pixel resolution of the model. The schematic for one jet is shown in  Figure \ref{schemes} (c).
The maser emission region begins 0.66 pc out along the jet axis and extends to the end. The  maser material can exist all
around the jet edge so it forms
a truncated hollow  cone, or it can be distributed through the jet, producing a solid cone.
These will be called the ``Hollow Jets'' and ``Solid Jets'' models.
If a  cross section is cut though a jet perpendicular to its axis producing a ring or disk, 
the flux over that ring or disk remains constant as it travels  out from the core with the jet material. The bi-conical jet structure can be  inclined in the same way as the jet carrying the ring described in the previous subsection. Since both jets are conical outflows we use the 
same method for obtaining the maser velocities as for the Grainy Ring, but with some differences. 
Since the maser material extends  along the jets we set the velocity to be decreasing linearly with distance outwards,
so the fastest material is near the core.
Because the velocity
range is centered around 0 km s$^{-1}$ with a red- and blue-shifted jet on either side, the line-of-sight velocity will be scaled to a range
of -50 to 50 km s$^{-1}$.
An example  of the Hollow Jets  inclined to the South-East by $i=30\,^{\circ}$ is shown in Figure \ref{schemes} (f).

\subsection{Spectrum and Source Path}
\label{variabilities}
We calculate the spectrum as the source travels  across a region of high magnification on the map, specifically
in a roughly S-E direction over one of the  N-S ``trunks''  of high magnification caustics visible in Figure \ref{big_map}.
This means the path will begin and end at areas of low magnification but pass across caustic regions of high density and complexity within the trunk. Such a path is a continuous
and representative sampling of the microlensing effects that are represented by the map. The path is chosen by eye from examining
the magnification maps at high resolution, and is slightly different for each magnification map used.   This method produces a time-varying spectrum from which two parameters are
extracted which are measures of variability:
\begin{enumerate}
\item A measure of the movement of the velocity centroid of the spectrum
\item The variation in total magnitude 
\end{enumerate}
In both cases the standard deviation of the distribution of values is used as the variability measure.

\section{Results and Discussion}
\label{Results}

\subsection{Time scale}
\label{time scale}

\hl{The magnification map can be used to find the duration of, and the time 
between, high magnification events, which occur when the source passes over regions of high magnification in the map.}
Using a Single Spot for the source, we extract a light curve spanning the total width from East to West of the map for
 image A1.  Scanning the light curve we find instances where the magnitude rises by 1, indicating the onset of an event. 
The shortest distance between two such instances is 0.034 pc, giving a time of 
195 years between events.  The shortest distance in which the magnitude rises by a value  1  is 0.003 pc, 
corresponding to a time of 17 years for the  rise time of an event.  
This translates into significant variability for sources with small substructure like spots.
For the spectrum of the Grainy Patch model, over a period of 17 years, 
statistical analysis shows that a  \hl{typical variation  of 0.12 mag and
2.0 km s$^{-1}$ in velocity centroid is possible.}
These paths with variability in short intervals are more likely to be
East-West
across the North-South pattern of the shear visible in the map.  A more detailed analysis is beyond this the scope of this paper but could be attempted in the future
when more information is available, and using more sophisticated techniques \citep[eg.][]{wyithe}.

\subsection{Model Variability Statistics}

\shl
The variability measures indicate how microlensing affects a source model as it moves behind the lens, i.e. along
the magnification map on the paths we have chosen.
A summary of the results is depicted in  the panels in Figure \ref{overall}. Each panel is a graph of a variability measure (vertical axis) as a function of source model (horizontal axis). The top two panels show the light curve and velocity centroid variation for all models lensed within image A1, the next two are the same, but for image A2. 
The horizontal axis is the same for all panels, and only listed on the bottom panel. The vertical axis may be magnitude (mag) or velocity
(km$s^{-1}$) for each panel, depending on the measure, and is labelled on the vertical axis of the panel.
The horizontal axis values are codes indicating the model and orientation. Major tick marks are labelled with a code indicating
the model: SS=Single Spot, SP=Smooth Patch, GR=Grainy Ring, SJ=Solid Jet, HJ=Hollow Jet. For the models that can
be oriented, this code is followed by a direction: S=South, \ehl\shl S-E=South East, E=East, and then $0^\circ$, indicating
the initial orientation of the model. The minor ticks \shlA that \ehlA follow are increasing values of inclination, only $30^\circ$ and 
$55\,^\circ$ are shown. The Grainy Ring is inclined by $0^\circ$, $10^\circ$, $20^\circ$, $30^\circ$, $30^\circ$, $50^\circ$, $55^\circ$, $60^\circ$ and $75^\circ$,
the Jets by the same, as well as $90^\circ$.    By scanning along the horizontal axis one can immediately \shlA see \ehlA the relative variability
of all the models, by the height of the graph in a panel.
\ehl

From these graphs several general comments can be made. The Grainy Patch and Ring produce the highest
variability in velocity centroid, both about the same value of 7 km s$^{-1}$, because they contain spots, i.e. small similar substructure. However the Grainy Patch produces more magnitude variability relative to the Ring because it is smaller overall than the Ring.
The Single Spot has no velocity centroid variation because its velocity range is below our resolution, but it has the highest magnitude variability (0.6), because it is the smallest-sized model. In contrast, the variability for the Jet models is low on both measures, because the models are large. The Smooth Patch has variability in both measures,  less than the Grainy models and more than the Jet models, and an unexpectedly high variation in velocity centroid, considering it has no substructure. This variation will be explained in section \ref{forms small} based on  its spectral behaviour.
Using a simple measure of  variability per model -- a sum of the rankings for centroid and magnitude for both images -- the 
models in descending order of variability are: Grainy Patch, 
Grainy Ring (best at i=$75\,^{\circ}$ East), 
Single Spot, Smooth Patch, 
Hollow Jets, Solid Jets. The overall variability is similar for both images A1 and A2.
\ehl 

The orientation relative to the shear changes the variability for those models that can be oriented. Any orientation that produces 
structures parallel to the direction of shear in  the map, i.e. N-S, will induce higher microlensing 
variability. This can be seen clearly in the centroid variability produced by the Grainy Ring in image A1
(Figure \ref{overall}, top panel).  As the direction changes from South to East, and with a high 
inclination, the ring presents a thin N-S profile, so the microlensing variability increases. In the case of the Jets, when they are N-S and highly inclined, the entire structure is in the direction of shear and so shows the most magnitude variability.
 
\subsection{The forms of the spectra}

The jets will be discussed separately from the other models since they have a different velocity spread and 
we will see they behave differently under inclinations.

\subsubsection{Spot, Patches and Ring}
\label{forms small}
The form of the spectra for these source models is given in Figure \ref{unlensed_spectra}. The Single Spot
shows a single peak with no velocity spread.  The Smooth Patch 
produces a Gaussian spectrum, as it is a Gaussian model. The Grainy Patch  has a multi-peaked spectrum due
to the many small spots. The Grainy Ring shows the typical spectrum shape generated by a ring structure, with
two peaks on the sides of the spectrum and a continuum between them, it is  slightly broken because of the grainy nature.
The spectra in Figure \ref{unlensed_spectra} are good representative examples as the spectra for all models retain those overall shapes even 
when inclined (remember the velocities have been normalised to always lie in the range of -350 to -250 km s$^{-1}$), and they
also retain that overall shape when lensed. The changes that occur due to microlensing can be described either as an
overall rise and fall or
the introduction/disappearance/movement/magnification  of  small peaks or folds. Figure \ref{lensed_spectra} contain
typical examples of these phenomena, each spectrum is the lensed spectrum over time corresponding to the unlensed spectrum of Figure \ref{unlensed_spectra}. 
The Single Spot shows how  the magnification can rise and drop significantly when the quasar is travelling across a trunk of high magnification on the magnification map. The peak in the velocity axis rises from Time = 0 yrs to a maximum when Time $\sim$ 500 yrs, and then drops sharply after that. The Smooth Patch when lensed has a similar peak that rises and falls over time, but in the Velocity axis the peak is wide and smooth. One
can also see that at Time=0 yrs, Velocity $\sim$ -280 km s$^{-1}$ there is a fold, which diminishes over time. Also,
although it is difficult to see in Figure \ref{lensed_spectra}(b),  the entire peak for the Smooth Patch drifts from left to right on the Velocity axis, explaining the high centroid variability for this model; otherwise, it would have a very low centroid variability. The Grainy Patch has many peaks which rise and fall at different times, due to the substructure in the model, with an overall rise and fall as in  the previous spectra.
The Grainy Ring varies in flux but only shows some  rise and fall at the edges, and at different times;   it also  shows an interior ridge formed by a peak that appears near Velocity $\sim$ -260  km s$^{-1}$  at Time $\sim$ 400 yrs
and then moves to the other side of the spectrum arriving at Velocity $\sim$ -340 km s$^{-1}$  at Time $\sim$ 700 yrs,
mimicking the drift in the Smooth Patch spectra. Other Ring orientations manifest similar behaviours.
 The rise and fall of the spectra, as in the Single Spot, will affect the total observed brightness of the maser
line over time; the change of folds and peaks will affect the location of the velocity centroid of the
broadened maser line over time. \shl We do not see any dramatic changes in spectrum shape, \ehl based on viewing all the spectra obtained. 

\subsubsection{Jets}

The jets are a bi-conical structure with identical outflowing  jets on either side. The velocity profile
is a linear velocity gradient out from the core, red- or blue-shifted on each side. With
this model the form of the unlensed spectrum can change significantly depending on the orientation, but not over time, in contrast
to the other models. Figure \ref{unlensed_spectra_jet}
shows two Solid Jets models  and two Hollow Jets models  before lensing. The Solid Jets and Hollow Jets when face-on
produce an identical spectrum because the fluxes are normalized to be the same, and  in this orientation the
observed velocities will be the same. The apparent peaks are due to the resolution of the velocity slicing, if the resolution 
was very small the spectra would be smooth and flat. 
When the same models are inclined by $50\,^{\circ}$, they
both transform into a single peak, which becomes sharper the more inclined they are.
Figure \ref{lensed_spectra_jet} shows 
the same models after lensing, and it is clear there is little change due to lensing, because the jet structure
is large compared to the Einstein Radius and does not undergo significant microlensing. There is some variation right at the peak with small
folds coming or going, but the flux is fairly constant, indicating little variation in the spectrum over time.  What is of more interest with this structure is the
spectra produced by different inclinations. Starting with a square wave when face-on, \emph{both} the
Solid and Hollow Jets spectra can merge towards a single smooth peak whether lensed or unlensed, with microlensing
producing slight variations at the peak over time. This has implications for matching
source models with the observed spectra.

\subsection{Comparison with observations}

Several factors make it difficult to choose which of our models can produce the observed maser line. \shl Firstly, in a source with multiple components, some components may be  de-magnified by microlensing and thus not appear
 in the spectrum; observation over long time scales may reveal these. Secondly,   
the entire maser spectrum may not be clearly visible
 because of  noise in the signal or components of the maser spectrum may lie below the detection limit. \ehl Thirdly, the observed line is coincident with an HI  absorption trough \citep{barvainis,curran}, possibly masking a different spectral signature.
Fourthly, the observed line is the combined spectra of both image A1 and A2. Although the images have similar microlensing properties and similar variability measures,   the  spatial  correlation  between the high resolution maps for
A1 and A2 is only 56\%. This indicates that some high magnification events do not correlate between
the two images in either magnification or epoch of observation. It
would be preferable to observe the maser line in A1 and A2 separately, since this will provide extra information to use to distinguish between models. Naturally the last three issues  are due to the technical limitations of the observation, and will improve in the future.

To match the observation, we look for models that have produced a single peak. The Single Spot has a peak, but it is not likely to be realistic since masers in jets usually occur in groups of spots. The Smooth Patch produces a smooth peak that is broad and maintains its shape, rising and falling with changing magnification, although it may drift around in velocity. Therefore if observation shows little  change in shape, but  changes in flux and centroid, the Smooth Patch is indicated. The Grainy Patch produces several peaks depending on the arrangement of maser spots, and our model deliberately includes a large peak which could match the observation. The existence of other smaller peaks in the model spectrum, but not seen in the observed spectrum, does not lessen the likelihood for  the Grainy Patch.
 Such peaks may be currently de-magnified or lie below the noise, either within the current observed line or at lower or higher velocities outside it. In the latter case the Grainy Patch model remains valid as our velocity range is artificial and can also be expanded. If future observations resolve smaller peaks, or peaks appear  over time,
the Grainy Patch model is consistent with the observation.
The Grainy Ring produces two clear peaks at the higher and lower ends of the spectrum and we do not
observe enough microlensing variation that could produce a single peak, so this model appears less likely. Interestingly,
the Jet models are capable of producing a smooth single peak, which varies insignificantly over time in either velocity or flux. If the observed
maser spectrum exhibited the same properties, then the Jet models
would be consistent with the observation. However, an inclination  \shl of more than 50$\,^{\circ}$  \ehl is required to produce such a spectrum, and since the quasar is believed to be not so inclined, this  mitigates against the Jet models. 

\subsection{Magnifications}
A search for masers in other lensed AGNs has not been successful  \citep{mckean}, partly due to the lack of required sensitivity in existing telescopes, but that will improve. However, predictions have been made about the detectability of lensed AGN masers  assuming a magnification of 35 for MG 0414+0534 \shl based on analytic strong lensing models
\citep{maser,trotter}.\ehl Using the Single Spot as a source, we found a peak magnification of 91 can be achieved in image A2, with
a magnification of 35 or better  achieved 22\% of the time.
\shl If long-term observing, combined with models of the system, show that the typical magnification is lower than 35, this indicates the maser is more intrinsically luminous than previously thought, with consequences for the rate of detection.
\ehl

\section{Conclusions}

Observations of jet masers in active galaxies indicate that the masers typically exist in a group with a velocity range of the order 100 km s$^{-1}$. The maser in MG 0414+0534 consists of a single peak. This is not consistent with
 a group of  maser spots that  produces multiple peaks, unless some peaks are lost in noise, or currently de-magnified, which is certainly possible. On the other hand the observation could be a smooth patch of masers, or a smooth distribution through the inner parts of the quasar jets, all of which also produce a single peak. 
A patch of spots will produce the most variability, in both velocity centroid and magnitude, the smooth patch will produce less -- it may  drift in velocity but remain fairly constant in shape -- the jets will not change at all.
A source with substructure like a patch of spots could produce changes in the spectrum within a couple of hundred years, lasting for 10-20 years.  These results are  just the beginning of the numerical analysis of this lensed maser, and 
continuing work would include:

\begin{itemize}
\item \shl Developing physically consistent models based on future observations and new understanding of masers
\item \shl Developing statistical tools for the extraction of  histograms, structure functions, and other data from the
high resolution magnification maps
\end{itemize}
but it would be \shl prudent to wait for   observations that have a higher sensitivity and resolution of flux and velocity,
and in particular that are able to detect the maser in images B and C. 
A long-term monitoring program \ehl should be carried out on this source, either with a large single dish or a sensitive VLBI array (e.g. the European VLBI Network) with higher  resolution than the original Effelsberg observations, with the following aims:
\begin{itemize}
\item monitoring should be conducted on monthly  time scales to pick up short term fluctuations
   as well as building statistics for long term variability
\item the A1 and A2 image spectra should be separated, which allow a comparison of the spectral behaviour between the two images for all the models, adding information that may constrain the choice of model
\item \sbl observations at higher spectral resolution may allow a more accurate determination of the spectral shape; differential fluctuations over time due to substructure will  help
   to determine the morphology of a maser group. \ehl
\end{itemize}

\section*{Acknowledgments}

Computing facilities were provided by the High Performance Computing Facility at The University of Sydney. 
This work is undertaken as part of the
Commonwealth Cosmology Initiative (www.thecci.org), and funded
 by the Australian Research Council  Discovery
 Project DP0665574.
We thank the anonymous referee whose comments improved the quality of this paper.

\bsp

\label{lastpage}

\end{document}